\documentclass[prb,twocolumn,eqsecnum,floatfix]{revtex4}

\usepackage{amsmath}
\usepackage{amsfonts}
\usepackage{epsfig}

\renewcommand{\vec}[1]{{\rm\bf #1}}
\newcommand{\Tr}{\mathop{\mathrm{Tr}}}

\renewcommand{\Im}{\mathop{\mathrm{Im}}}
\renewcommand{\Re}{\mathop{\mathrm{Re}}}
\newcommand{\sign}{\mathop{\mathrm{sgn}}}
\newcommand{\Ei}{\mathop{\mathrm{Ei}}}
\newcommand{\vep}{\varepsilon}
\newcommand{\ep}{\epsilon}

\newcommand{\unitmatrix}{\openone}
\newcommand{\phonon}{\mu}
\newcommand{\lep}{\lambdabar_\ep}
\newcommand{\omph}{\omega_\mathrm{ph}}
\newcommand{\ellph}{\ell_\mathrm{ph}}
\newcommand{\valley}{\kappa}

\begin{document}

\title{Boundary problems for Dirac electrons and edge-assisted Raman scattering in graphene}
\author{D.~M.~Basko}\email{denis.basko@grenoble.cnrs.fr}
\affiliation{Laboratoire de Physique et Mod\'elisation des Mileux Condens\'es,
Universit\'e Joseph Fourier and CNRS,
25 Rue des Martyrs, BP 166, 38042 Grenoble, France}
\begin{abstract}
The paper reports a theoretical study of scattering of electrons by edges
in graphene and its effect on Raman scattering. First, we discuss effective
models for translationally invariant and rough edges. Second, we employ these
models in the calculation of the edge-activated Raman $D$~peak intensity and
its dependence on the polarization of the incident/scattered light, as well
as on the position of the excitation spot. Manifestations of the quasiclassical
character of electron motion in Raman scattering are discussed.
\end{abstract}

\maketitle

\section{Introduction}


Graphene, a monolayer of carbon atoms, first obtained in
2004,\cite{Novoselov2004} received increasing interest due to its
exceptional electronic properties and original transport physics.%
\cite{Novoselov2007}
Gradual miniaturization of graphene devices increases the importance
of edge effects with respect to the bulk two-dimensional physics.
Starting from a graphene sheet, nanoribbons and quantum dots can be
produced by lithography and etching.%
\cite{Ruoff1999,Kim2007,Avouris2007,Novoselov2008,Ensslin2009,Goldhaber2009}
Edges also play a fundamental role in the quantum Hall effect.

Graphene edges can be studied by several experimental techniques.
Scanning tunneling microscopy (STM) and transmission elecron
microscopy (TEM) can resolve the structure of the edge on the
atomic scale.%
\cite{Klusek2000,Kobayashi2005,Kobayashi2006,Dresselhaus2008,Liu2009,Ritter2009}
Raman scattering has also proven to be a powerful technique
to probe graphene edges.%
\cite{Cancado2004,Novotny,You2008,Gupta2009,Casiraghi2009}
The so-called $D$~peak at 1350~cm$^{-1}$ is forbidden by
momentum conservation in a perfect infinite graphene crystal,
and can only be activated by impurities or edges. Invoking
the double-resonance mechanism for the $D$~peak activation,%
\cite{ThomsenReich2000} Can\c{c}ado \emph{et al.} have
shown that a perfect zigzag edge does not give rise to the
$D$~peak.\cite{Cancado2004}
It should be emphasized that this property is determined by
the effect of the edge on the electronic states.

A great deal of theoretical studies of electronic properties
near the edge has focused on the case of ideal zigzag or
armchair edges, most commonly adopting the tight-binding
description. One of the spectacular results obtained by
this approach was the existence of electronic states confined
to the zigzag edge,%
\cite{Stein1987,Tanaka1987,Fujita1996-1,Fujita1996-2,Dresselhaus1996}
which was later confirmed experimentally.%
\cite{Klusek2000,Kobayashi2005,Kobayashi2006,Ritter2009}
The question about general boundary condition for Dirac electron
wave function at a translationally invariant graphene edge has
been addressed\cite{McCann2004} and a detailed analysis of
boundary conditions which can arise in the tight-binding model
has been performed.\cite{Akhmerov2008}
In spite of the fact that all graphene samples produced so
far have rough edges, the number of theoretical works dedicated
to rough edges is limited. Most of them model edge roughness
in the tight-binding model by randomly removing lattice sites.%
\cite{Areshkin2007,Guinea2007,Querlioz2008,Evaldsson2008}
The opposit limit of smooth and weak roughness has been
considered.\cite{Fang2008}
Edge states on zigzag segments of finite length have also 
been studied recently.\cite{Tkachov2009}

The present work has several purposes.
One is to develop analytically treatable models which would
describe electron scattering on various types of edges in terms
of as few parameters as possible.
The second one is to calculate the polarization dependence of
the $D$~peak intensity for different models of the edge, and
thus see what information can be extracted from this dependence.
The third one is to identify the characteristic length scale
which confines the Raman process to the vicinity of the edge,
i.~e. the spatial extent of the Raman process.
It will be shown that the last two issues are intimately related
to the quasiclassical character of the electron motion during
the Raman scattering process.

The paper is organized as follows. In Sec.~\ref{sec:Qualitative}
we discuss the problem in qualitative terms and summarize the
main results of the work. In Sec.~\ref{sec:Free} we summarize
the Dirac description of single-electron states in an infinite
graphene crystal and formulate the Huygens-Fresnel principle
for Dirac electrons.
In Sec.~\ref{sec:Edge} we discuss models for the electron
scattering from a graphene edge, considering translationally
invariant as well as rough edges.
Sec.~\ref{sec:Phonons} introduces the model for electron-phonon
coupling and describes the general scheme of the calculation of
the $D$~peak intensity using the standard perturbation theory
in the coordinate representation. Finally,
Secs.~\ref{sec:Regular}, \ref{sec:Rough}, and
\ref{sec:fragmented} are dedicated to the calculation of the
$D$~peak intensity for an ideal armchair edge, an atomically
rough edge, and an edge consisting of a random collection of
long zigzag and armchair segments, respectively.

\section{Qualitative discussion and summary of the main results}
\label{sec:Qualitative}

\subsection{Electron scattering by the edge}
\label{sec:Qreflection}

\begin{figure}
\includegraphics[width=8cm]{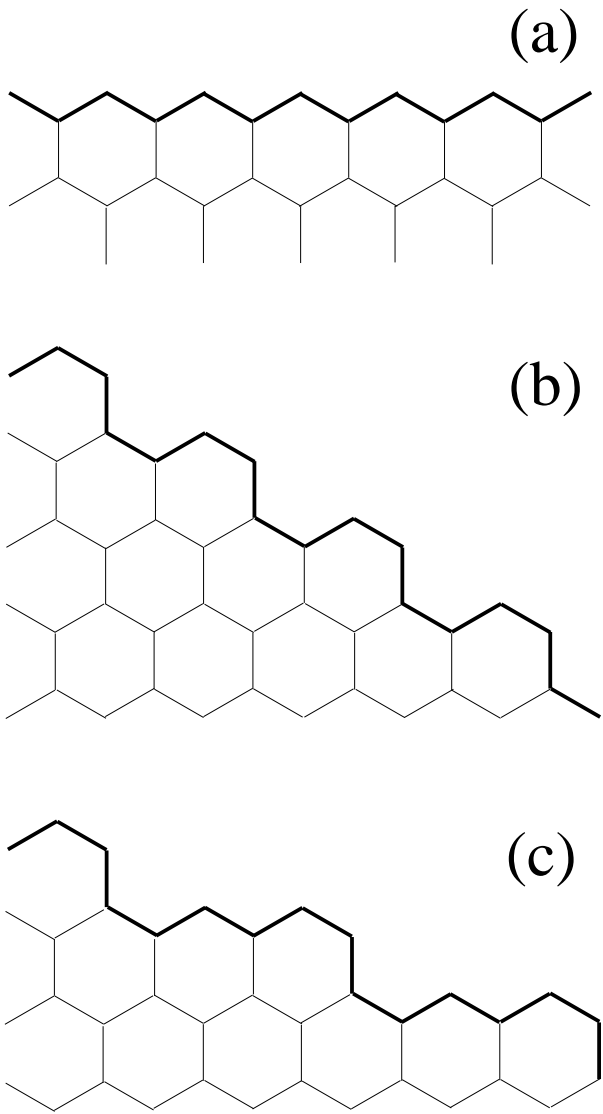}
\caption{\label{fig:edges} Examples of ordered edges: (a)~zigzag,
(b)~armchair, (c)~a more complicated but still translationally
invariant edge.
}
\end{figure}

First, we discuss translationally invariant edges. For example
(see Fig.~\ref{fig:edges}), a zigzag edge has a spatial period
$d_e=a\sqrt{3}$ ($a\approx{1}.42\:\mbox{\AA}$ is the C--C bond
length), an armchair edge has $d_e=3a$, and a more complicated
edge, shown in Fig.~\ref{fig:edges}(c), has 
$d_e=\sqrt{21}\,a\approx{4.6}\,a$ (the spatial period is
measured along the average direction of the edge).
It is important to compare
$d_e$ to the electronic wavelength (we prefer to divide the
latter by $2\pi$), $\lep\equiv{v}/|\ep|$, where $\ep$~is the
electron energy, and $v\approx{1}.1\cdot{10}^8~\mbox{cm/s}\approx%
{7}.3\:\mbox{eV}\cdot\mbox{\AA}$ is the electron velocity (the
slope of the Dirac cones). For comparison, at $\ep=1\:\mbox{eV}$
$\lep\approx 7.3\:\mbox{\AA}\approx{5}a$.
As long as $d_e<\pi\lep$, the component of the electronic momentum
along the edge, $p_\|$, is conserved (we measure the electron
momentum from the Dirac point). For longer periods the edge acts
analogously to a reflective diffraction grating in optics; this
case is not considered here.
In the limit $d_e\ll\lep$ the reflection of electrons from any
periodic edge can be described by an effective energy-independent
boundary condition for the electronic wave
function.\cite{McCann2004,Akhmerov2008}

Next, we study rough edges. An extreme case is when the edge is
rough at the atomic scale, like that in the tight-binding model
with randomly removed sites. Then it is reasonable to assume
that in the vicinity of the edge all plane-wave components with
a given energy in both valleys are mixed randomly, as there is
no small or large parameter which would suppress or favor any
particular channel (the smallness $a/\lep\ll{1}$ suppresses the
direct intravalley scattering, but multiple intervalley scattering
efficiently mixes states within the same valley as well). The
electron is thus scattered randomly both in all directions (as
a consequence of the randomization of the momentum direction
within the same valley), and between the two valleys. For this
case in Sec.~\ref{sec:irregular} we
propose a phenomenological model which describes such random
scattering of electrons by the edge, respecting only the
particle conservation and the time-reversal symmetry.
Essentially, each point of the edge is represented by an
independent point scatterer which randomly rotates the valley
state of the electron. This model is used for quantitative
calculations in the subsequent sections.

Edges, rough on length scales much larger than the lattice
constant, are likely to consist of distinct segments of zigzag
and armchair edges, as shown by
STM\cite{Kobayashi2005,Kobayashi2006,Ritter2009} and
TEM.\cite{Dresselhaus2008,Liu2009}
Then the overall probability of scattering within the same valley
or into the other valley is simply determined by the fraction of
the corresponding segments. The problem of angular distribution
of the scattered electrons is analogous to the well-studied
problem of light scattering by rough surfaces.%
\cite{Brown1984,Maradudin1985a,Maradudin1985b,GarciaStoll,%
TranCelli,Maradudin1989,Maradudin1990}
The main qualitative features of the scattering, namely, the
sharp coherent peak in the specular direction, the smooth
diffuse background, and the enhanced backscattering peak,
should be analogous for the electrons in graphene as well.
The so-called surface polaritons, shown to play an important
role in the light scattering, are analogous to the electronic
edge states in graphene. Still, full adaptation of this theory
for the case of Dirac electrons in graphene represents a
separate problem and is beyond the scope of the present work.
Here we only consider the case when regular edge segments are
sufficiently long, i.~e. their typical length $d_e\gg\lep$.
Then the diffraction corrections, small in the parameter
$\lep/d_e\ll{1}$, can be simply found as it is done in the
classical optics,\cite{BornWolf} using the Huygens-Fresnel
principle for Dirac electrons, Eq.~(\ref{Huygens=}).

\subsection{Quasiclassical picture of Raman scattering}
\label{sec:quasiclassical}

Since graphene is a non-polar crystal, Raman scattering involves
electronic excitations as intermediate states: the electromagnetic
field of the incident laser beam interacts primarily with the electronic
subsystem, and emission of phonons occurs due to electron-phonon
interaction. The matrix element of the one-phonon Ramanprocess
can be schematically represented as
\begin{equation}\label{Ramanmatrixelement=}
\mathcal{M}\sim\sum_{a,b}
\frac{\langle{i}|\hat{H}_{e-em}|a\rangle\langle{a}|\hat{H}_{e-ph}|b\rangle
\langle{b}|\hat{H}_{e-em}|f\rangle}
{(E_i-E_a+2i\gamma)(E_i-E_b+2i\gamma)}.
\end{equation}
Here $|i\rangle$ is the initial state of the process (the incident
photon with a given frequency and polarization, and no excitations
in the crystal), $|f\rangle$ is the final state (the scattered photon
and a phonon left in the crystal), while $|a\rangle$ and $|b\rangle$
are the intermediate states where no photons are present, but an
electron-hole pair is created in the crystal and zero phonons or one
phonon have been emitted, respectively. Note that these intermediate
states correspond to electronic eigenstates in the presence of the
edge, i.~.e, scattered states rather than plane waves.
$E_i=E_f$, $E_a$, and~$E_b$ are the energies of the corresponding
states, and $2\gamma$ is inverse inelastic scattering time (the
overall rate of phonon emission and electron-electron collisions).
$\hat{H}_{e-em}$~and~$\hat{H}_{e-ph}$ stand for the
terms in the system hamiltonian describing interaction of electrons
with the electromagnetic field and with phonons, respectively.

As discussed in Refs.~\onlinecite{shortraman,megapaper}, for
one-phonon scattering processes it is impossible to satisfy the
energy conservation in all elementary processes. This means that
the electron-hole pair represents a virtual intermediate state,
and no real populations are produced. Formally, at least one of
the denominators in Eq.~(\ref{Ramanmatrixelement=}) must be at
least of the order of the phonon frequency~$\omph\gg\gamma$. In
fact, the main contribution to the matrix element comes from such
states that the electron and the hole have the energy $\ep$ close
(within $\sim\omph$) to half of the energy $\omega_{in}$ of the
incident photon: $|\ep-\omega_{in}/2|\sim\omph$. These two energy
scales are well separated: $\omph\approx{0}.17$~eV, while typically
$\omega_{in}/2\approx{1}$~eV. According to the uncertainty principle,
the energy uncertainty, $\omph$, determines the typical lifetime of
the virtual state (electron-hole pair), $\sim{1}/\omph$. This time
scale determines the duration of the whole process.

As we are dealing with a translationally non-invariant system,
it is useful to analyze the Raman process in the coordinate
representation. The time scale $1/\omph$, introduced above,
translates into the length scale $\ellph=v/\omph$.
Thus, this length scale,
$\ellph\approx{4}\:\mbox{nm}$, determines the spatial extent of
the process (we will return to this point below). Its largness
compared to the electron wavelength,
$\ellph/\lep=\omega_{in}/(2\omph)\gg{1}$, ensures that the
electronic wave functions determining the matrix elements for
each elementary process, {\em admit a quasiclassical
representation}. The quasiclassical approximation for the
electronic wave functions is fully analogous to the geometrical
optics approximation for electromagnetic waves, electronic
trajectories corresponding to light rays. Corrections to
this approximation are known as diffraction and are small by the
parameter $\omph/\omega_{in}\ll{1}$. It should be emphasized that
the quasiclassical picture is neither an assumption, nor a
hypothesis, but it arises automatically in the direct calculation
of the Raman matrix element which is performed in the main part
of the paper.

\begin{figure}
\centerline{\includegraphics[width=8cm]{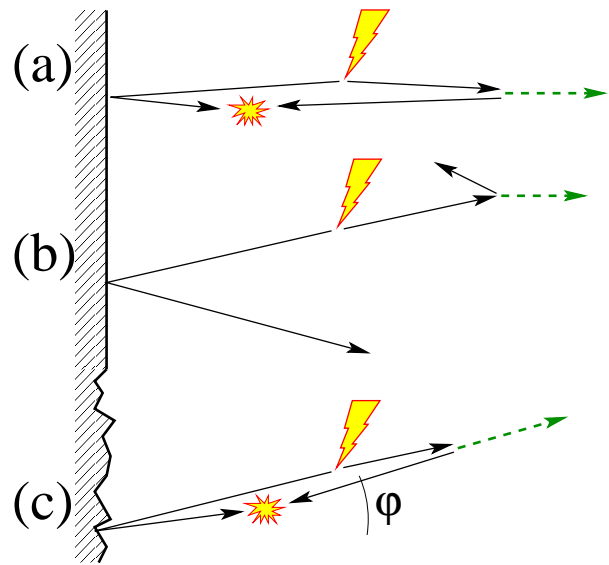}}
\caption{\label{fig:backscatt} (Color on-line) Real-space
representation of the scattering process responsible for the
D-peak near graphene edges. The lightning represents the
incoming photon which generates the electron-hole pair. The
solid black arrows represent the quasi-classical trajectories
of the electron and the hole. The dashed arrow represents the
emitted phonon. The flash represents the radiative recombination
of the electron-hole pair producing the scattered photon.
(a)~Backscattering off a translationally invariant edge is
possible only at normal incidence (up to the quantum uncertainty).
(b)~For oblique incidence on a translationally invariant edge
the reflection is specular, so the electron and the hole will
not be able to meet at the same point.
(c)~For a rough edge backscattering is possible even at
oblique incidence.
}\end{figure}

In the quasiclassical picture, the photoexcited electron and hole
can be viewed as wave packets of the size $\sim\lep$, initially
created at an arbitrary point of the sample. More precisely,
instead of a point one can consider a region of a size $\delta{l}$,
such that $\lep\ll\delta{l}\ll\ellph$. Then momentum conservation
holds up to $\delta{p}\sim{1}/\delta{l}\ll\epsilon/v$ by virtue
of the uncertainty principle, so that electron and hole momenta
whose magnitude is $\epsilon/v$ (counted from the Dirac point),
have approximately opposite directions, as the photon momentum
is very small. The same argument holds for the phonon emission
and for the radiative recombination process: in order to emit a
photon, the electron and the hole must meet in the same region of
space of the size~$\delta{l}$ with almost opposite momenta (up to
$1/\delta{l}$). Momentum conservation at the reflection from the
edge depends on the quality of the edge, as discussed in
Sec.~\ref{sec:Qreflection}. Regardless of the properties of the
edge, an elementary geometric consideration, illustrated by
Fig.~\ref{fig:backscatt}, shows that for the electron and the hole
to be able to meet, the scattering on both the phonon and on the
edge must be backward.

\begin{figure}
\includegraphics[width=8cm]{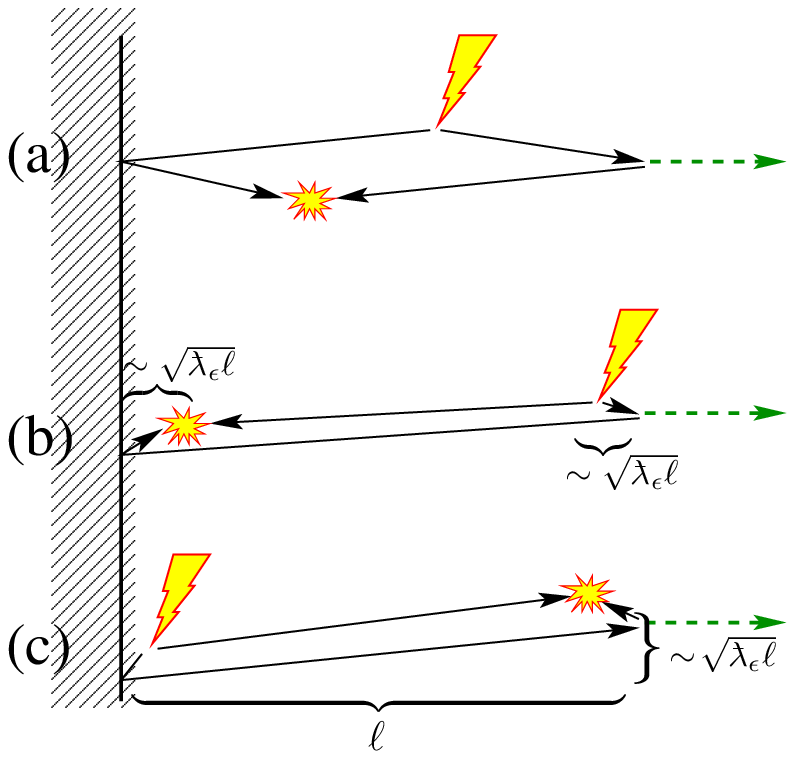}
\caption{\label{fig:traject}(Color on-line.) Real space representation
of different contributions to the matrix element of the scattering
process responsible for the D-peak at an ideal armchair edge, placed
at $x=0$.
The solid black arrows represent the
quasi-classical trajectories of the electron and the hole corresponding
to the three Green's functions in Eqs.~(\ref{D1tr=}),~(\ref{D2tr=}).
Trajectories (a), (b), (c) correspond to decomposition of each of the
three Green's functions in Eq.~(\ref{D1tr=}).
$\ell$~is the overall spatial extent of the process, and
$\lep=v/\ep$ is the electron wavelength divided by 2$\pi$.
}
\end{figure}

\begin{figure}
\includegraphics[width=8cm]{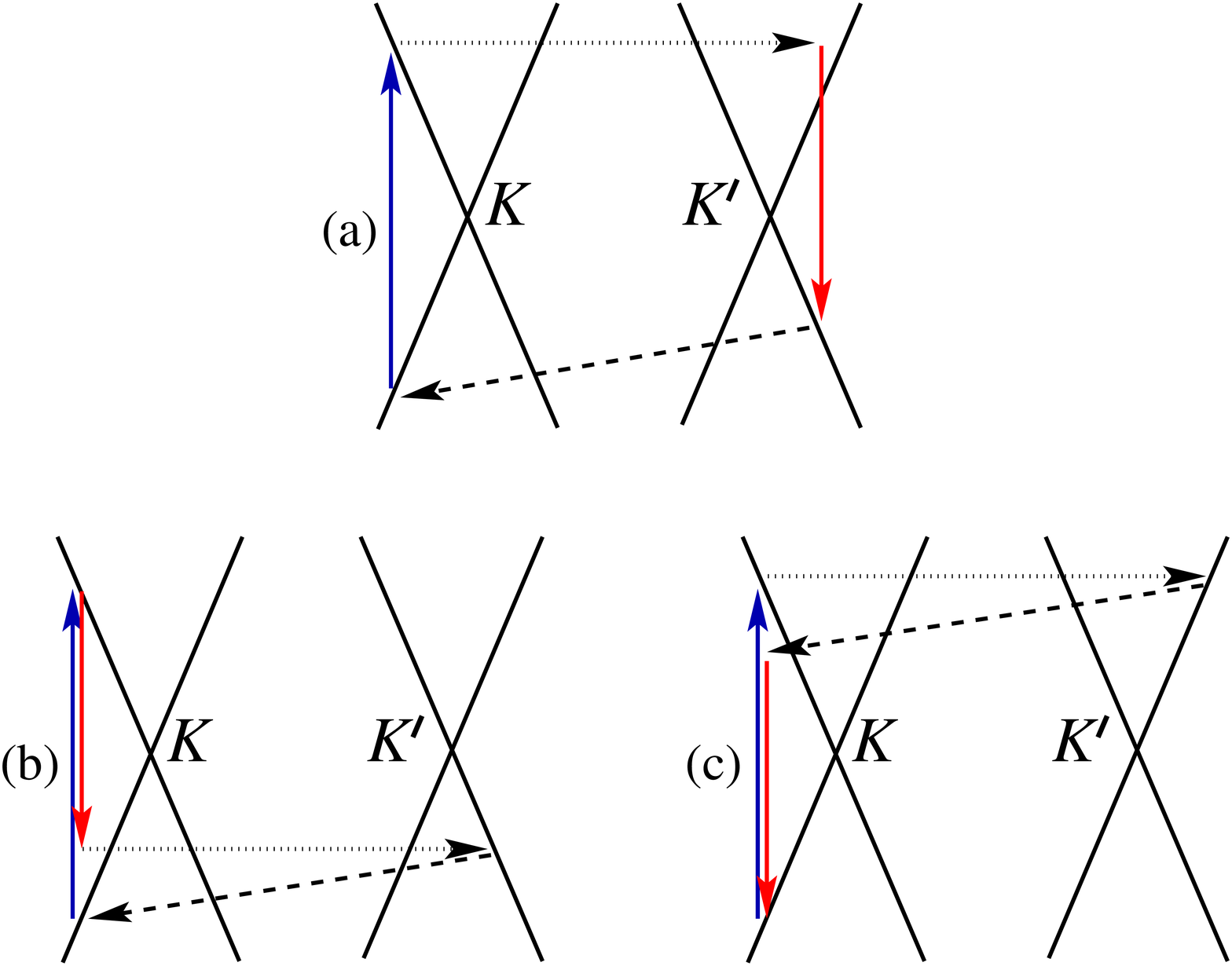}
\caption{\label{fig:resfig}(Color on-line.) Momentum space representation
of different contributions to the matrix element of the scattering
process responsible for the D-peak at an armchair graphene edge,
corresponding to the real space picture shown in Fig.~\ref{fig:traject}.
Solid lines represent the Dirac cones around $K$~and~$K'$ points of the
first Brillouin zone.
Vertical solid arrows represent interband electronic transitions accompanied
by photon absorption or emission (photon wave vector is neglected), dashed
arrows represent phonon emission, the horisontal dotted arrow represents
the scattering from the edge.
}
\end{figure}

In the quasiclassical picture, the electron and the hole have to travel the
same distance between creation and annihilation, as their velocities are
equal. Then the process in Fig.~\ref{fig:traject}~(a) has more phase space
satisfying this restriction, and this process gives the main contribution to
the Raman matrix element. This will be also shown by an explicit estimate
in Sec.~\ref{sec:integrated}. Note that the three processes shown in
Fig.~\ref{fig:traject}, can be considered in the momentum space, as shown
in Fig.~\ref{fig:resfig}. According to the abovesaid, the processes
(b)~and~(c), often shown in the literature as an illustration of the
double resonance,\cite{ThomsenReich2000} are in fact weaker than the
process~(a) by a factor $\sim\omph/\omega_{in}$.

\subsection{Polarization dependence}

The backscattering condition has immediate consequences for the
polarization dependence of the Raman scattering intensity. Indeed,
the matrix element of creation/annihilation of an electron and a
hole pair with momenta $\vec{p},-\vec{p}$ (counted from the Dirac
point) by a photon with the polarization~$\vec{e}$, is proportional
to $[\vec{e}\times\vec{p}]_z$, reaching its maximum when
$\vec{e}\perp\vec{p}$. Since a perfect edge conserves the
component of momentum along the edge, the backscattering is possible
only at normal incidence, as seen from Fig.~\ref{fig:backscatt}).%
\cite{singularity}
This gives the polarization dependence of the D peak intensity as
$I_D\propto\sin^2\varphi_{in}\sin^2\varphi_{out}$, where
$\varphi_{in}$ and $\varphi_{out}$ are the angles between the
polarizations of the incident and scattered photons and the normal
to the edge.
If one does not use the analyzer to fix the polarization of the
scattered photons, the dependence is $I_D\propto\sin^2\varphi_{in}$.
In experiments, however, the intensity never goes exactly to zero
for the polarizations perpendicular to the edge, but remains a
finite fraction~$\varepsilon$ of the intensity in the maximum.%
\cite{Cancado2004,Gupta2009,Casiraghi2009}
What determines this residual intensity in the minimum? 

For an ideal edge the finite value of the intensity in the
minimum is entirely due to the \emph{quantum uncertainty}.
Namely, the momenta of the electron and the hole upon their
creation are not exactly opposite, but up to an uncertainty
$\sim{1}/\delta{l}$; the annihilation occurs not exactly at
the same spatial point, but within the spatial
uncertainty~$\delta{l}$. If the spatial extent of the process
is~$\ell$, the uncertaintly is estimated as
$\delta{l}\sim\sqrt{\lep\ell}$, and the ratio $\varepsilon$
of the intensities in the minimum and in the maximum (i.~e.,
for polarizations perpendicular and parallel to the edge,
respectively) should be small as a power of the small
parameter quasiclassical parameter $\lep/\ell$. The calculation
is performed in Sec.~\ref{sec:integrated}, the result is given
by Eqs.~(\ref{IDregularb=}) and (\ref{IDregularc=}) for the
detection without and with the analyzer, respectively. Up
to logaritmic factors, the ratio
$\varepsilon\sim\omph^2/\omega_{in}^2\sim(\lep/\ellph)^2$.
This corresponds to $\ellph\sim\ell$, in accordance with
the energy-time uncertainty principle, as discussed in the
previous subsection.

For rough edges the intensity in the minimum is determined by
the ability of the edge to backscatter electrons at oblique
incidence, as shown in Fig.~\ref{fig:backscatt}(c).
If the edge is rough at the atomic scale, oblique
backscattering is nearly as efficient as normal backscattering.
Still, such oblique trajectories are longer than those
corresponding to the normal incidence, so they are expected to
have a smaller weight since the virtual electron-hole pair
lives only for a restricted time. So one still can expect a
minimum of intensity for perpendicular polarization, but it is
of a purely geometrical origin, so one does not expect a
parametric smallness of the ratio~$\varepsilon$.
The calculation using the model Sec.~\ref{sec:irregular} for an
atomically rough, is performed in Sec.~\ref{sec:disordpolariz},
and the result is given by
Eqs.~(\ref{IDdisordb=}) and (\ref{IDdisordc=}) for the
detection without and with the analyzer, respectively.
In the former case the ratio $\varepsilon=1/3$, and the minimum
is indeed for the polarization of the incident light perpendicular
to the edge. With the analyzer, the absolute minimum is
$\varepsilon={1}/10$, reached when the polarizer and the analyzer
are oriented at the angle $\pi/3$ with respect to the edge and to
each other. When the polarizer and the analyzer are both rotated
parallel to each other, the minimum is $\varepsilon=1/5$.

For an edge consisiting of segments longer than electronic
wavelength, $d_e\gg\lep$, we first analyze the contribution of
a single armchair segment. It is calculated in
Sec.~\ref{sec:fragmented}, and given by Eqs.~(\ref{IDfragma=})
and (\ref{IDfragmb=}) for the detection without and with the
analyzer, respectively.
The minimum is reached for the polarization, perpendicular
to the \emph{armchair direction} (which does not have to
coincide with the average direction of the edge!), and is
determined by the \emph{quantum diffraction} of the electron
on the segment, $\varepsilon\sim\lep/d_e$ (provided that
$\lep/d_e\gtrsim\omph^2/\omega_{in}^2$, the ratio for the
infinite armchair edge, which is the case when
$d_e\lesssim{50}\:\mbox{nm}$ for $\omega_{in}=2\:\mbox{eV}$).
To obtain contribution of the whole
edge, it is sufficient to multiply these expression by the
total number of such segments and replace $d_e$ by its average,
provided that all armchair segments have the same orientation.

It is crucial, however, that up to three different orientations
of armchair segments are possible, at the angle $\pi/3$ to each
other. When contributions corresponding to several different
orientations are added together, the polarization dependence
may change quite dramatically,
as was briefly discussed by the author and coworkers in
Ref.~\onlinecite{Casiraghi2009} and is considered in more
detail in Sec.~\ref{sec:fragmented}.
Note that if the average direction of the edge is armchair or
zigzag, the possible orientations of armchair segments are
symmetric with respect to the average direction: three
orientetaions at angles 0 and $\pm\pi/3$ for an armchair
edge, and two at angles $\pm\pi/6$ for a zigzag edge. Most
likely, the number of segments corresponding to ``$+$''
and ``$-$'' signs will be equal on the average. Then, by
symmetry, the maximum of intensity in both cases will be
reached for the polarization along the average direction of
the edge, in agreement with recent experimental
observations.\cite{Gupta2009,Casiraghi2009}

When the average direction is armchair, the ratio
between the minimum and the maximum of intensity is
determined by the relative fraction of segments oriented
at $\pm\pi/3$ with respect to those oriented along the
average direction.
When the average direction is zigzag, the polarization
dependence is fully determined by the symmetry (on the
average) between the two armchair directions, and is
given by Eq.~(\ref{IDzigzag=}). The ratio between
the minimum and the maximum is $\vep=1/3$ for detection
without analyzer, and $\vep=1/9$ for detection with an
analyzer parallel to the polarizer of the incident light,
again, in agreement with
Refs.~\onlinecite{Gupta2009,Casiraghi2009}.
Thus, quantum diffraction effects appear to be masked
by the purely geometrical effects.

Remarkably, the quantum diffraction limit is still
accessible if only \emph{two} orientations of armchair
segments are present (which is the case when the average
direction is zigzag or close to it). It is sufficient
to put the polarizer perpendicular to one of the armchair
directions, and the analyzer perpendicular to the other
one, thereby killing the leading specular contribution
for both segments. In this polarization configuration the
absolute minimum of the intensity is reached, and it is
indeed determined by the quantum diffraction, as given
by Eq.~(\ref{IDquantum=}).

\subsection{Excitation position dependence}
\label{sec:qualpos}

In Ref.~\onlinecite{Novotny}, confocal Raman spectroscopy was
suggested and used as a way to probe the length scale~$\ell$
which restricts the $D$ peak to be in the vicinity of the edge
(the spatial extent of the Raman process).
The idea is to focus the incident light beam as tightly as
possible, so that its electric field $\mathcal{E}_{in}(\vec{r})$
has the shape
$\mathcal{E}_{in}(\vec{r})\propto{e}^{-|\vec{r}-\vec{r}_0|^2/(2L^2)}$,
where the width $L$ can be measured independently, and the spot
center position~$\vec{r}_0$ can be varied experimentally.
Then, given that the intensity of the $D$~peak is proportional to
\begin{equation}
I_D\propto\int\mathcal{K}(\vec{r},\vec{r}')\,
\mathcal{E}_{in}(\vec{r})\,\mathcal{E}_{in}^*(\vec{r}')\,
d^2\vec{r}\,d^2\vec{r}',
\end{equation}
where $\mathcal{K}(\vec{r},\vec{r}')$ is a certain kernel,
decaying away from the edge with a characteristic length
scale~$\ell$, a measurement of the dependence $I_D(\vec{r}_0)$
would give information on the kernel.
The measurement would be especialy simple in the case $L\ll\ell$.
In reality, however, the relation is the opposite:
in Ref.~\onlinecite{Novotny} $L=186.5\:\mbox{nm}$, and $\ell$ is
a few tens of nanometers at most.\cite{citeGupta}

In this situation the
dependence of the Raman intensity $I_D(\vec{r}_0)$ is very close
to the excitation intensity profile $|\mathcal{E}_{in}(\vec{r})|^2$,
and the nonlocality of the kernel $\mathcal{K}(\vec{r},\vec{r}')$
manifests itself only in a slight change of shape of$I_D(\vec{r}_0)$
with respect to $|\mathcal{E}_{in}(\vec{r})|^2$. In the first
approximation it can be viewed just as small shift and broadening.
When the signal-to-noise ratio is not sufficiently high to perform
the full functional deconvolution, one has to assume a specific
functional form for the kernel and do a few-parameter fit. It is
clear that different functional forms will give values of~$\ell$,
differing by a factor of the order of~1.
In Ref.~\onlinecite{Novotny} the form
$\mathcal{K}(\vec{r},\vec{r}')=%
\theta(x)\,\theta(x')\,e^{-2\gamma(x+x')/\ell_\gamma}$
was assumed, where $x$ is the distance from the edge, $\theta(x)$
is the step function, and $\ell_\gamma=v/(2\gamma)$ is the electron
inelastic scattering length [$2\gamma$ is the electron inelastic
scattering rate, see Eq.~(\ref{Ramanmatrixelement=})]. This
assumption seems to contradict the fact that the lifetime of the
virtual electron-hole pair is $\sim{1}/\omph$, discussed in
Sec.~\ref{sec:quasiclassical}, as it was pointed out in
Refs.~\onlinecite{megapaper,Casiraghi2009}.

The explicit form of the kernel $\mathcal{K}(\vec{r},\vec{r}')$
for an ideal armchair edge is calculated in Sec.~\ref{sec:regSpatial},
it is given by Eq.~(\ref{spatialkernel=}), and it turns out to be more
complicated than a simple exponential. In fact, it depends on both
length scales, $\ellph$ and $\ell_\gamma$. The length $\ellph$ is
shorter, but the spatial cutoff it provides is only power-law. The
longer length~$\ell_\gamma$ is responsible for the strong exponential
cutoff.
Which of the two lengths plays the dominant role in the Raman process,
turns out to depend on the {\em specific observable} to be measured.
The total integrated intensity for
$\mathcal{E}_{in}(\vec{r})=\mathrm{const}$
is proportional to the integral
$\int\mathcal{K}(\vec{r},\vec{r}')\,d^2\vec{r}\,d^2\vec{r}'$,
which is determined mainly by $\ellph$, while $\ell_\gamma$ enters
only in a logarithmic factor. The same can be said about
the polarization dependence and diffraction corrections, discussed
in the previous subsection. However, the change in the shape of
$I_D(\vec{r}_0)$, compared to the excitation intensity profile
$|\mathcal{E}_{in}(\vec{r})|^2$, is determined by the second and
higher \emph{moments} of the kernel,
$\int{x}^n(x')^{n'}\,\mathcal{K}(\vec{r},\vec{r}')\,%
d^2\vec{r}\,d^2\vec{r}'$, with $n+n'\geq{2}$. These moments turn
out to be determined by the longer scale $\ell_\gamma$. Thus, the
interpretation by the authors of Ref.~\onlinecite{Novotny} of their
experiment is qualitatively correct.

Analysis of the experimental data of Ref.~\onlinecite{Novotny}
using kernel~(\ref{spatialkernel=}) gives
$\ell_\gamma=66\:\mbox{nm}$, corresponding to
$2\gamma\approx{11}\:\mbox{meV}$. 
Analogous analysis was done for the case of strongly disordered
edge in Sec.~\ref{sec:disordspatial}, and in this model one
obtains $\ell_\gamma=120\:\mbox{nm}$. Indeed, as the disordered
edge gives more weight to oblique trajectories, as shown in
Fig.~\ref{fig:backscatt}(c), the effective distance from the
edge at which the kernel decays, is shorter than for the
normal incidence, so a larger value of $\ell_\gamma$ is required
to fit the data.

The inelastic scattering rate for an electron with the
energy~$\ep$ due to phonon emission can be written as
$2\gamma=(\lambda_\Gamma+\lambda_K)\ep/2$,\cite{shortraman,megapaper}
where $\lambda_\Gamma$ and $\lambda_K$ are dimensionless
electron-phonon coupling constants [$\lambda_K$ is defined in
Eq.~(\ref{lambdaphonon=}), and $\lambda_\Gamma$~is defined
analogously, but the optical phonons at the $\Gamma$~point should
be considered].
The value of the constant~$\lambda_\Gamma$ can be reliably
taken to be about $\lambda_\Gamma\approx{0}.03$.
Indeed, a DFT calculation\cite{Piscanec2004} gives
$\lambda_\Gamma\approx{0}.028$; measurements of the dependence
of the $G$-peak frequency $\omega_G$ on the electronic Fermi
energy~$\ep_F$,
$d\omega_G/d|\epsilon_F|\approx\lambda_\Gamma/2\pi$ give
$\lambda_{\Gamma}\approx{0}.034$\cite{Yan2007} and
$\lambda_{\Gamma}\approx{0}.027$;\cite{Pisana2007}
the value of $\lambda_\Gamma$ is not renormalized by the Coulomb
interaction.\cite{BaskoAleiner}
The value of $\lambda_K$ has been debated recently.%
\cite{Calandra2007,BaskoAleiner,Lazzeri2008}
The measurements of the phonon group velocity (see
Ref.~\onlinecite{Lazzeri2008} for the summary of the experimental
data) give $\lambda_K\approx{0}.04$.
The ratio between the two coupling constants can be also
extracted from the experimental ratio of the two-phonon
peak intensities,\cite{megapaper}
$2(\lambda_K/\lambda_\Gamma)^2\approx{20}$,\cite{Ferrari2006}
which gives $\lambda_K\approx{0}.10$.

Thus, $\lambda_\Gamma+\lambda_K\approx{0}.1\pm{0.03}$, seems
to be a reasonable estimate. This estimate gives
$2\gamma\approx{50}\:\mbox{meV}$ for electrons with the
energy $\omega_{in}/2=0.98\:\mbox{eV}$, which translates
into a value of $\ell_\gamma$ several times shorter than
that following from the results of Ref.~\onlinecite{Novotny}.

\subsection{On the Tuinstra-K\"onig relation}

For a sample of a finite size $L_a$ the total $D$~peak intensity is
proportional to the total length of the edge, i.~e., the sample
perimeter, so that $I_D\propto{L}_a$.
At the same time, the intensity of the $G$~peak at 1581~cm$^{-1}$
is propotional to the area of the sample, i.~e., $I_G\propto{L}_a^2$.
These simple facts result in the so-called Tuinstra-K\"onig
relation, established in experiments on graphite nanocrystallites
long ago,\cite{Tuinstra1970,Knight1989} and confirmed experimentally
many times afterwards:%
\cite{Dresselhaus1999,Cancado2006,Cancado2007,Sato2006}
$I_D/I_G\propto{1}/L_a$.
The proportionality coefficient cannot be determined universally;
it clearly depends on the character of the edge 
(as an extreme case, one can imagine a hexagonal flake with entirely
zigzag edges which do not give any $D$~peak at all, except at the
junctions between them; then $I_D$ is not even proportional
to $L_a$). 

What is the boundary of validity of the Tuinstra-K\"onig relation
on the small-size side?
At the same time, it was noted in Ref.~\onlinecite{Ferrari2000}
that since the atomic dispacement pattern corresponding to the $D$~peak
must involve at least one aromatic ring, the size of the ring, a few
angstroms, represents an absolute lower bound. From the results of
the present work it follows that the dependence $I_D\propto{L}_a$
becomes logarithmically sensitive to the presence of the opposite
edge, $I_D\propto{L}_a\ln(\omph{L}_a/v)$ if the size is smaller the
electron inelastic length $L_a<v/(2\gamma)$, and the whole approach
becomes invalid when $L_a\sim{v}/\omph\approx{4}\:\mbox{nm}$.
The breakdown of the $1/L_a$ dependence has indeed been observed
for $L_a$ smaller than a few nanometers.\cite{Zickler2006}

\section{Free Dirac electrons}\label{sec:Free}

\begin{figure}
\includegraphics[width=8cm]{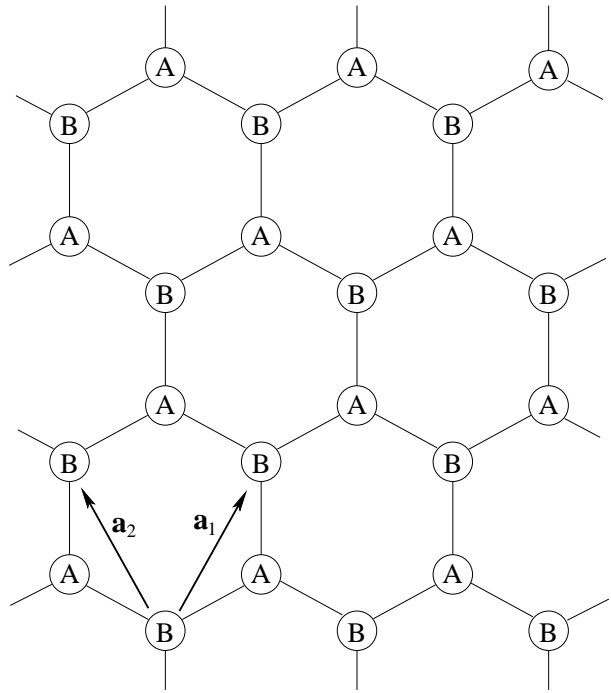}
\caption{\label{fig:lattice} Honeycomb lattice with the $A$ and $B$
sublattices and the elementarty translation vectors.
}
\end{figure}

\begin{table}
\begin{tabular}[t]{|c|c|c|c|c|c|c|} \hline
$C_{6v}$ & $E$ & $C_2$ & $2C_3$ & $2C_6$ & $\sigma_{a,b,c}$ &
$\sigma_{a,b,c}'$
\\ \hline\hline $A_1$ & 1 & 1 & 1 & 1 & 1 & 1 \\ \hline $A_2$ & 1
& 1 & 1 & 1 & $-1$ & $-1$ \\ \hline $B_2$ & 1 & $-1$ & 1 & $-1$ &
$1$ & $-1$ \\ \hline $B_1$ & 1 & $-1$ & 1 & $-1$ & $-1$ & $1$ \\
\hline $E_1$ & 2 & $-2$ & $-1$ & $1$ & 0 & 0 \\ \hline $E_2$ & 2 & 2 &
$-1$ & $-1$ & 0 & 0 \\ \hline\end{tabular}\hspace{1cm}
\caption{Irreducible representations of the group $C_{6v}$ and their
characters.\label{tab:C6vC3v}}
\end{table}

\begin{table}
\begin{tabular}{|c|c|c|c|c|c|c|} \hline
irrep & 
$A_1$ & $B_1$ & $A_2$ & $B_2$ & $E_1$ & $E_2$ \\ \hline
\multicolumn{7}{|c|}{valley-diagonal matrices}\\
\hline matrix & $\unitmatrix$ & $\Lambda_z$ & $\Sigma_z$ &
$\Lambda_z\Sigma_z$ & $\Sigma_x,\,\Sigma_y$ &
$-\Lambda_z\Sigma_y,\Lambda_z\Sigma_x$ \\ \hline
\multicolumn{7}{|c|}{valley-off-diagonal matrices} \\
\hline matrix & $\Lambda_x\Sigma_z$ & $\Lambda_y\Sigma_z$ & $\Lambda_x$
& $\Lambda_y$ & $\Lambda_x\Sigma_y,-\Lambda_x\Sigma_x$ &
$\Lambda_y\Sigma_x,\Lambda_y\Sigma_y$ \\ \hline
\end{tabular}
\caption{Classification of $4\times{4}$ hermitian matrices according to
irreducible representations (irreps) of the $C_{6v}$~group.
\label{tab:matrices}}
\end{table}

In this section we summarize the model for the bulk graphene only,
which is fully analogous to that of Ref.~\onlinecite{megapaper}.
Properties of the edge are discussed in the next section.

We measure the single-electron energy~$\ep$ from the Fermi level of the
undoped (half-filled) graphene. The Fermi surface of undoped graphene
consists of two points, called $K$~and~$K'$. Graphene unit cell
contains two atoms, labeled $A$ and $B$ (see Fig.~\ref{fig:lattice}),
each of them has one $\pi$-orbital, so there are
two electronic states for each point of the first Brillouin zone (we
disregard the electron spin). Thus, there are exactly four electronic
states with zero energy. An arbitrary linear combination of them is
represented by a 4-component column vector~$\psi$. States with low
energy are obtained by including a smooth position dependence
$\psi(\vec{r})$, $\vec{r}\equiv(x,y)$. The low-energy hamiltonian has
the Dirac form:
\begin{equation}\label{Hel=}
\hat{H}_0=\int{d}^2\vec{r}\,\hat\psi^\dagger(\vec{r})\,
(-iv\vec\Sigma\cdot\vec\nabla)\,\hat\psi(\vec{r}).
\end{equation}
Here we used the second-quantized notation and introduced
the electronic $\psi$-operators $\hat\psi(\vec{r}),\hat\psi^\dagger(\vec{r})$.

It is convenient to define the $4\times{4}$ isospin matrices
$\vec\Sigma\equiv(\Sigma_x,\Sigma_y)$, not through their explicit
form, which depends on the choice of the basis (specific arrangement
of the components in the column~$\psi$), but through their
transformation properties. Namely, all 16 generators of the $SU(4)$
group, forming the basis in the space of $4\times{4}$ hermitian
matrices, can be classified according to the irreducible
representations of~$C_{6v}$, the point group of the graphene crystal
(Tables \ref{tab:C6vC3v} and~\ref{tab:matrices}). They can be
represented as products of two mutually commuting algebras of Pauli
matrices $\Sigma_x,\Sigma_y,\Sigma_z$ and
$\Lambda_x,\Lambda_y,\Lambda_z$,\cite{McCann2006,AleinerEfetov} which
fixes their algebraic relations. By definition, $\Sigma_x,\Sigma_y$ are
the matrices, diagonal in the $K,K'$ subspace, and transforming
according to the $E_1$~representation of~$C_{6v}$.

In the following we will take advantage of the symmetry with
respect to time reversal. The action of the time-reversal operation
on the four-component envelope function $\psi(\vec{r})$ is
defined as
\begin{equation}\label{UTpsi=}
\psi(\vec{r})\mapsto
U_t\psi^*(\vec{r}),
\end{equation}
where $U_t$~is a unitary $4\times{4}$ matrix.  when
applied twice, the time reversal operation should give an identity,
which results in an additional requirement $U_tU_t^*=\unitmatrix$.
The explicit form of $U_t$ depends on the choice of the basis.

In those rare cases when a specific representation has to be chosen,
we use that of Ref.~\onlinecite{AleinerEfetov},
\begin{equation}\label{Aleinerrep=}
\psi=\left[\begin{array}{c}
\psi_{AK} \\ \psi_{BK} \\ \psi_{BK'} \\ -\psi_{AK'}
\end{array}\right],
\end{equation}
where the first subscript labels the sublattice ($A,B$), and the
second one labels the valley ($K,K'$). In this basis $\Sigma_i$~are
the Pauli matrices acting within upper and lower 2-blocks of the
column (the sublattice subspace), while $\Lambda_i$ are the Pauli
matrices acting in the ``external'' subspace of the 2-blocks (the
valley subspace).  The time reversal matrix in this representation
is given by $U_t=\Sigma_y\Lambda_y$.

The electron Green's
function, corresponding to the hamiltonian~(\ref{Hel=}), is given by
\begin{equation}\label{Greenfp=}
G_0(\vec{p},\ep)=
\frac{\ep+v\vec{p}\cdot\vec\Sigma}{\ep^2-(vp-i\gamma_\ep)^2},
\end{equation}
where $\vec{p}$ and $\ep$ are electronic momentum and energy, counted
from the Dirac point. The inelastic broadening $\gamma_\ep\ll|\ep|$ is
introduced phenomenologically. In the coordinate representation
the Green's function is given by
\begin{eqnarray}
G_0(\vec{r},\ep)
&=&\frac{\ep+i\gamma_\ep\sign\ep-iv\vec\Sigma\cdot\vec\nabla}{4iv^2}\,H_0^{(1)}(\zeta),
\label{Greenfr=}
\end{eqnarray}
where $H_0^{(1)}(\zeta)$ is the Hankel function and
$\zeta\equiv(|\ep|+i\gamma_\ep)r/v$.
We will mostly need the asymptotic form valid at distances
$r\gg|\ep|/v$,
\begin{eqnarray}
G_0(\vec{r},\ep)=-\sqrt{\frac{i\zeta}{2\pi}}\,\frac{e^{i\zeta}}{vr}
\left[\frac{\sign\ep+\vec\Sigma\cdot\vec{r}/r}{2}
\left(1+\frac{i}{8\zeta}\right)\right.\nonumber\\
-\left.\frac{\sign\ep-\vec\Sigma\cdot\vec{r}/r}{2}\,\frac{i}{4\zeta}+O(\zeta^{-2})\right].
\label{GreenfLarger=}
\end{eqnarray}

Any wave function $\psi(\vec{r})$ satisfying the Dirac equation,
$(\ep+i\gamma_\ep\sign\ep+iv\vec\Sigma\cdot\vec\nabla)\psi(\vec{r})=0$,
in some region~$\mathcal{O}$ of space, satisfies also the
Huygens-Fresnel principle. Namely, the value of $\psi(\vec{r})$
at an arbitrary point $\vec{r}\in\mathcal{O}$ can be written as
an integral over the boundary~$\partial\mathcal{O}$,
\begin{equation}
\psi(\vec{r})=iv\oint\limits_{\partial\mathcal{O}}
\vec{n}\cdot\left[G_0(\vec{r}-\vec{r}_e,\ep)\,
\vec\Sigma\,\psi(\vec{r}_e)\right]d\vec{r}_e.
\label{Huygens=}
\end{equation}
Here $\vec{r}_e$ is the distance along the boundary,
$\vec{n}$ is the inner normal to the boundary.
This relation follows from the Gauss theorem and the fact that
$(p^2+\nabla^2)H_0^{(1)}(pr)=4i\delta(\vec{r})$ for any~$p$.

\section{Models for electrons near the edge}\label{sec:Edge}

\subsection{Translationally invariant edge}\label{sec:edgeReg}

The main assumption of this subsection is that the component of
the electronic momentum~$\vec{p}$ along the edge is conserved upon
reflection, so that a plane wave is reflected as a plane wave. The
most studied ideal zigzag and armchair edges fall into this category.
Here we do not restrict ourselves just to zigzag or armchair edges,
requiring only that the spatial period~$d_e$ of the edge is smaller
than half the electron wavelength, $d_e<\pi\lep$.
For $d_e\ll\lep$ the reflection of electrons from the edge can be
described by an effective boundary condition for the electronic wave
function.\cite{McCann2004,Akhmerov2008}

The edge is assumed to be a straight line determined by its normal
unit vector~$\vec{n}$,  so that graphene occupies the half-plane
$\vec{n}\cdot\vec{r}>0$.
The microscopic Schr\"odinger equation determines the effective
boundary condition on the wave function~$\psi(\vec{r})$, which
for smooth functions (on the scale~$d_e$) can be simply written as
$\left.B\psi\right|_\mathrm{edge}=0$, where $B$~is a $4\times{4}$
hermitian matrix.  The rank of~$B$ is equal to~2 since the linear
space of incident states at fixed energy is two-dimensional due
to the valley degeneracy. Thus, it has two zero eigenvalues, while
the other two can be set to~1 without the loss of generality (only
the zero subspace of~$B$ matters). Thus, one can impose the
condition $B^2=B$. Equivalently, one can write $B=(\unitmatrix-M)/2$,
where $M$~has the same eigenvectors as~$B$, but its eigenvalues are
equal to $\pm{1}$, hence $M^2=\unitmatrix$. To ensure current
conservation, the condition $\left.B\psi\right|_\mathrm{edge}=0$
must automatically yield $\psi^\dagger(\vec{n}\cdot\vec\Sigma)\psi=0$;
this means that $M(\vec{n}\cdot\vec\Sigma)+(\vec{n}\cdot\vec\Sigma)M=0$.
Finally, the time reversal symmetry requires that the conditions
$B\psi=0$ and $BU_t\psi^*=0$ must be equivalent, which yields
$M^*=U_t^\dagger{M}U_t$.
To summarize all the above arguments, the general energy-independent
boundary condition has the form
\begin{equation}
\left.(\unitmatrix-M)\psi\right|_\mathrm{edge}=0,
\label{boundarycond=}
\end{equation}
where the $4\times{4}$ hermitian matrix~$M$ satisfies the following
conditions, which result to be the same as obtained in Ref.~\onlinecite{McCann2004}:
\begin{equation}
M^2=\unitmatrix,\quad
M(\vec{n}\cdot\vec\Sigma)+(\vec{n}\cdot\vec\Sigma)M=0,\quad
M=U_tM^*U_t^\dagger.
\end{equation}
Matrices satisfying these constraints can be parametrized
by an angle~$\chi$ and a three-dimensional unit vector $(m_x,m_y,m_z)$:%
\cite{Akhmerov2008}
\begin{subequations}\begin{eqnarray}
M&=&\left(\Sigma_z\cos\chi+[\vec{n}\times\vec\Sigma]_z\sin\chi\right)
M_\Lambda,\label{edgeMa=}\\
M_\Lambda&=&\sum_{i=x,y,z}m_i\Lambda_i,\;\;\;
m_i\in\mathbb{R},\;\;\;
\sum_{i=x,y,z}m_i^2=1.\label{edgeMb=}
\end{eqnarray}\label{edgeM=}\end{subequations}
Without loss of generality we can assume $\cos\chi\geq{0}$
(the negative sign can always be incorporated in $M_\Lambda$).

Explicit expressions for the matrix~$M$, corresponding
to the edges shown in Fig.~\ref{fig:edges}, can be obtained
in the tight-binding model on the terminated honeycomb
lattice\cite{Brey2006} [it is convenient to use
representation~(\ref{Aleinerrep=})].
For the zigzag edge with $\vec{n}=-\vec{e}_y$
[Fig.~\ref{fig:edges}(a)] the boundary condition is
$\psi_{BK}=\psi_{BK'}=0$, which gives $M=\Sigma_z\Lambda_z$.
This agrees with the prediction of Ref.~\onlinecite{Cancado2004}
that upon reflection from a zigzag edge the electron cannot
scatter between the valleys, which corresponds to a valley-diagonal
matrix $M$.
For the armchair edge with $\vec{n}=\vec{e}_x$
[Fig.~\ref{fig:edges}(a) rotated by $2\pi/3$ counterclockwise]
we have $\psi_{AK}+\psi_{AK'}=\psi_{BK}+\psi_{BK'}=0$, and
$M=-\Sigma_y\Lambda_y$. It can be shown that in
the nearest-neigbor tight-binding model on a terminated
honeycomb lattice only zigzag and armchair boundary
conditions can be obtained, the latter occurring only if
the edge direction is armchair, while to obtain the full
form of Eq. (\ref{edgeMa=}), one has to include an on-site
potential in the vicinity of the edge.\cite{Akhmerov2008}

It is known for quite some time that a perfect zigzag edge supports
states, confined to the edge.%
\cite{Stein1987,Tanaka1987,Fujita1996-1,Fujita1996-2,Dresselhaus1996}
Let us see what class of boundary conditions is compatible with existence
of such states. The wave function of an edge state must have the form
\begin{equation}
\psi(\vec{r})=\psi_0e^{-\kappa(\vec{n}\cdot\vec{r})+ip_\|[\vec{n}\times\vec{r}]_z},
\quad\kappa>0,
\end{equation}
where the vector $\psi_0$ is such that the solution satisfies both the Dirac
equation in the bulk with some energy~$\ep$, as well as the boundary
condition at the edge.
It is convenient to make a unitary substitution
$\psi_0=e^{i\Sigma_x\chi/2}e^{-i\Sigma_z\varphi_\vec{n}/2}\tilde\psi_0$,
where $\varphi_\vec{n}$ is the polar angle of the direction~$\vec{n}$.
Then the two conditions have the following form:
\begin{subequations}\begin{eqnarray}
&&\left(i\kappa\Sigma_x+p_\|\cos\chi\Sigma_y+p_\|\sin\chi\Sigma_z\right)\tilde\psi_0
=\frac\ep{v}\,\tilde\psi_0,\\
&&\Sigma_zM_\Lambda\tilde\psi_0=\tilde\psi_0.
\end{eqnarray}\end{subequations}
The boundary condition is satisfied by two linearly independent vectors
$\tilde\psi_\pm$ which can be chosen to satisfy
$\Sigma_z\tilde\psi_\pm=\pm\tilde\psi_\pm$
(to find them it is sufficient to diagonalize the matrix~$M_\Lambda$).
Each of them satisfies the first condition if and only if
$\ep=\pm{v}p_\|\sin\chi$, $\kappa=\mp{p}_\|\cos\chi$.
The requirement $\kappa>0$ leaves only one of them,
$\tilde\psi_{-\sign{p}_\|}$, and the energy of the edge state is
$\ep=-v|p_\||\sin\chi$. Thus, it seems that almost any edge
can support a bound state, the exception being the case
$\cos\chi=0$ (armchair-like edge), which thus seems to be a special
rather than a general case.

Now we turn to the scattering (reflecting) states, which are the ones
responsible for the edge-assisted Raman scattering.
Even though the component~$p_\|$ of the electron momentum~$\vec{p}$,
parallel to the edge, is conserved, reflection can change the valley structure
of the electron wave function. The general form of such solution is
\begin{eqnarray}
&&\psi(\vec{r})=
\psi_\vec{p}
e^{ip_\perp(\vec{n}\cdot\vec{r})+ip_\|[\vec{n}\times\vec{r}]_z}+
\nonumber\\&&\qquad{}+
S_\Lambda[\vec{n}\times\vec\Sigma]_z
\psi_\vec{p}
e^{-ip_\perp(\vec{n}\cdot\vec{r})+ip_\|[\vec{n}\times\vec{r}]_z},
\label{reflectingsolution=}
\end{eqnarray}
where
$p_\perp=(\vec{n}\cdot\vec{p})<0$, $p_\|=[\vec{n}\times\vec{p}]_z$,
and $\psi_\vec{p}$ is an eigenvector of
$(\vec{p}\cdot\vec\Sigma)\psi_\vec{p}=\pm|\vec{p}|\psi_\vec{p}$.
The first term represents the wave incident on the edge, the second one
is the reflected wave. The matrix $[\vec{n}\times\vec\Sigma]_z$ simply
aligns the isospin of the reflected particle with the new direction of
momentum. The unitary matrix~$S_\Lambda$ represents a rotation in the valley
subspace. It should be found from the boundary condition~(\ref{boundarycond=})
(this is conveniently done in the basis of the eigenvectors of~$M_\Lambda$),
which gives
\begin{subequations}\begin{eqnarray}
&&S_\Lambda=\zeta\,
\frac{\Lambda_0+M_\Lambda(\cos\chi+\zeta^*\sin\chi)}%
{\Lambda_0+M_\Lambda(\cos\chi+\zeta\sin\chi)},\\
&&\zeta=\pm\frac{1}p
\left\{-[\vec{n}\times\vec{p}]_z+i(\vec{n}\cdot\vec{p})\right\},\quad
|\zeta|=1.
\end{eqnarray}\end{subequations}
For $\sin\chi=0$ (zigzag edge) we have $S_\Lambda=\zeta$.
For $\cos\chi=0$ (armchair edge) we have $S_\Lambda=M_\Lambda$,
independent of the direction of~$\vec{p}$.
The reflected part of Eq.~(\ref{reflectingsolution=}) can be
identically rewritten using the Huygens-Fresnel principle,
Eq.~(\ref{Huygens=}), as
\begin{equation}
\psi(\vec{r})=\psi_\vec{p}e^{i\vec{p}\vec{r}}
-\int\limits_\mathrm{edge} d\vec{r}_e\,
G_0(\vec{r}-\vec{r}_e,\ep)\,
v\Sigma_zS_\Lambda
\psi_\vec{p}e^{i\vec{p}\vec{r}_e},\label{reflsolHuyg=}
\end{equation}
so that $-v\Sigma_zS_\Lambda$ can be viewed as the $T$-matix
of the edge.

When $S_\Lambda$ does not depend on the direction of~$\vec{p}$
(armchair edge), it is easy to write down the exact explicit
expression for the single-particle Green's function:
\begin{eqnarray}\nonumber
G(\vec{r},\vec{r}';\ep)&=&G_0(\vec{r}-\vec{r}',\ep)\nonumber\\
&&{}+ G_0(\vec{r}-\vec{r}'+2\vec{n}(\vec{n}\cdot\vec{r}'),\ep)\,
[\vec{n}\times\vec\Sigma]_zS_\Lambda.\nonumber\\ && \label{imagesource=}
\end{eqnarray}
The second term represents nothing but the contribution of a fictitious
image source of particles, appropriately rotated, and placed at the
point $\vec{r}'-2\vec{n}(\vec{n}\cdot\vec{r}')$ obtained
from~$\vec{r}'$ by the reflection with respect to the edge.
In the quasiclassical approximation (analogous to geometric optics),
Eq.~(\ref{imagesource=}) is also valid for a general edge at large
distances $r,r'\gg{v}/|\ep|$, provided that position-dependent $S_\Lambda$
is taken, determined by
\begin{equation}
\zeta=-\sign\ep\,\frac{i(\vec{n}\cdot(\vec{r}+\vec{r}'))+
[\vec{n}\times(\vec{r}-\vec{r}')]_z}
{\sqrt{(\vec{n}\cdot(\vec{r}+\vec{r}'))^2
+[\vec{n}\times(\vec{r}-\vec{r}')]_z^2}}.
\end{equation}
Again, using the Huygens-Fresnel principle, we can rewrite
Eq.~(\ref{imagesource=}) identically as
\begin{eqnarray}\nonumber
&&G(\vec{r},\vec{r}';\ep)=G_0(\vec{r}-\vec{r}',\ep)\nonumber\\
&&\qquad{}-\int\limits_\mathrm{edge} d\vec{r}_e\,G_0(\vec{r}-\vec{r}_e,\ep)\,
v\Sigma_zS_\Lambda\,G_0(\vec{r}_e-\vec{r}',\ep).\nonumber\\
\label{HuygensG=}
\end{eqnarray}

\subsection{Atomically rough edge}
\label{sec:irregular}

As discussed in Sec.~\ref{sec:Qreflection}, when the edge is rough
on the atomic length scale, electron scattering is random both
in all directions and between the two valleys.
This case will be of main interest for us, as it (i)~represents
the opposite limiting case to that of an ordered edge, and (ii)~can
be described by a simple model proposed below.
The main assumption is that each point of an atomically rough edge 
acts as an independent point scatterer, independent from other
segments.
Associating thus a $T$-matrix to each point of the edge, we write
the scattered wave function in the form
\begin{eqnarray}
&&\psi(\vec{r})=\psi_\vec{p}e^{i\vec{p}\vec{r}}\nonumber\\&&\qquad{}+
\int\limits_{\mathrm{edge}}d\vec{r}_e\,
G_0(\vec{r}-\vec{r}_e,\ep)\,
T(\vec{s},\vec{e}_\vec{p};\vec{r}_e)\,
\psi_\vec{p}e^{i\vec{p}\vec{r}_e},\label{wfTmatrix=}\\
&&\vec{s}=\frac{\vec{r}-\vec{r}_e}{|\vec{r}-\vec{r}_e|},\quad
\vec{e}_\vec{p}=\frac{\vec{p}}{|\vec{p}|},\nonumber
\end{eqnarray}
where energy argument of the Green's function $\ep=\pm{v}p$ for
electrons and holes, respectively. The (one-dimensional)
integration is performed along the edge, which is assumed to be
a straight line determined, as in the previous subsection, by the
condition $(\vec{n}\cdot\vec{r}_e)=0$, where the unit vector
$\vec{n}$ is the normal to the edge. The unit vectors
$\vec{e}_\vec{p}$ and $\vec{s}$ indicate the incident and
scattering directions.

The $T$-matrix must satisfy
(i)~the particle conservation condition (unitarity), and
(ii)~the time reversal symmetry (reciprocity),
\begin{equation}
T(\vec{s},\vec{e}_\vec{p};\vec{r}_e)
=U_t\,T^T(-\vec{e}_\vec{p},-\vec{s};\vec{r}_e)\,U_t^\dagger.\label{Ttimerev=}
\end{equation}
Here $T^T$ stands for the $4\times{4}$ matrix transpose.
Unitarity and reciprocity are discussed in Appendix~\ref{app:Smatrix}
in the context of a general scattering theory, similar to
that for light scattering on a rough surface.\cite{Brown1984}
We propose the following form of the $T$-matrix:
\begin{subequations}
\begin{equation}\label{Tmatrix=}
T(p\vec{s},\vec{p};\vec{r}_e)=-\sqrt{\rho(\vec{s},\vec{e}_\vec{p})}\,
v\Sigma_zS_\Lambda(\vec{r}_e).
\end{equation}
The angular factor $\rho(\vec{s},\vec{e}_\vec{p})$ ensures the particle
conservation (unitarity of the edge scattering, see
Appendix~\ref{app:Smatrix} for details),
\begin{equation}\label{rho=}
\rho(\vec{s},\vec{e}_\vec{p})=
\frac{-2(\vec{n}\cdot\vec{e}_\vec{p})(\vec{n}\cdot\vec{s})}%
{1-(\vec{n}\cdot\vec{e}_\vec{p})(\vec{n}\cdot\vec{s})
-[\vec{n}\times\vec{e}_\vec{p}]_z[\vec{n}\times\vec{s}]_z}.
\end{equation}
If we introduce the  angles of incidence and scattering by
$(\vec{n}\cdot\vec{e}_\vec{p})=-\cos\varphi_i$,
$(\vec{n}\cdot\vec{s})=\cos\varphi_s$,
$[\vec{n}\times\vec{e}_\vec{p}]_z=\sin\varphi_i$,
$[\vec{n}\times\vec{s}]_z=\sin\varphi_s$, the specular
direction corresponding to $\varphi_s=\varphi_i$,
then
$\rho=2\cos\varphi_i\cos\varphi_s/[1+\cos(\varphi_i+\varphi_s)]$.
Note that since the structure of the wave functions in the
$\Sigma$-subspace is fixed by the direction of momentum, one may
suggest slightly different forms of Eq.~(\ref{rho=}) which would be
equivalent. For example, the $T$-matrix obtained by from
Eq.~(\ref{rho=}) by replacement of $\Sigma_z$ by
$i[\vec{n}\times\vec\Sigma]_z$ and by changing the sign of the third
term in the denominator in Eq.~(\ref{rho=}) would have the same matrix
elements between Dirac eigenstates of the same energy.

For a coordinate-independent $S_\Lambda$ the $\vec{r}_e$ integration
eliminates all directions $\vec{s}$ different from the specular one.
Since for the latter $\rho=1$, Eq.~(\ref{Tmatrix=}) reduces to
Eq.~(\ref{reflsolHuyg=}). In this case $S_\Lambda$ can be identified
with the scattering matrix. For the short-range disorder such
identification is not possible, since scattering process is necessarily
non-local on the length scale of at least $\sim{1}/p$. The connection
between the $T$-matrix in Eq.~(\ref{wfTmatrix=}) and the scattering
matrix is more complicated, and is discuss in detail in
Appendix~\ref{app:Smatrix}. Here we only mention that it would make
absolutely no sense to require the unitarity of $S_\Lambda(\vec{r}_e)$
at each given point. Instead, we use the following form:
\begin{equation}
S_\Lambda(\vec{r}_e)=\varpi(\vec{r}_e)\sum_{i=x,y,z}m_i(\vec{r}_e)\,\Lambda_i,
\quad \sum_{i=x,y,z}m_i^2(\vec{r}_e)=1.\label{SLambdaCOE=}
\end{equation}
Here $\varpi(\vec{r}_e)$ is a complex gaussian random variable whose
real and imaginary parts are distributed independently and identically
(so its phase is uniformly distributed between 0 and $2\pi$). The
numbers $m_x,m_y,m_z$, which can be viewed as components of a unit
three-dimensional vector, must be real to ensure the time-reversal
symmetry. One may assume them to be constant or taking just few
definite values, which would correspond the edge to be composed of
segments with $a\ll{d}_e\lesssim{1}/p$ of definite types (e.~g., zigzag
or armchair). For $d_e\sim{a}$, when the scattering between the valleys
is completely random, the vector $(m_x,m_y,m_z)$ can be taken uniformly
distributed over the unit sphere. We assume the matrices
$S_\Lambda(\vec{r}_e)$ to be uncorrelated at different points on the
edge, distant by more than~$d_e$, by writing
\begin{equation}
\overline{\varpi(\vec{r}_e)\,\varpi^*(\vec{r}_e')}=
\frac{\pi{v}}{|\epsilon|}\,\delta(\vec{r}_e-\vec{r}_e').
\label{COEaverage=}
\end{equation}
\end{subequations}
Here the overline denotes the ensemble averaging. However, we assume
that this product is self-averaging upon spatial integration, i.~e.,
that Eq.~(\ref{COEaverage=}) holds even in the absence of the ensemble
averaging when integrated over a sufficiently long segment of the edge
(namely, longer than~$d_e$). The prefactor $\pi{v}/|\ep|$ at the
$\delta$-function ensures the unitarity of scattering, see
Appendix~\ref{app:Smatrix}.

Eq.~(\ref{wfTmatrix=}) for the wave function yields an analogous
expression for the Green's function, valid sufficiently far from
the edge, $(\vec{n}\cdot\vec{r})\gg{v}/|\ep|$,
$(\vec{n}\cdot\vec{r}')\gg{v}/|\ep|$:
\begin{eqnarray}
&&G(\vec{r},\vec{r}';\ep)=G_0(\vec{r}-\vec{r}',\ep)+\nonumber\\
&&\qquad{}+\int\limits_{\mathrm{edge}}d\vec{r}_e\,
G_0(\vec{r}-\vec{r}_e,\ep)\,T(\vec{s},\vec{s}';\vec{r}_e)\,
G_0(\vec{r}_e-\vec{r}',\ep),
\label{gfTmatrix=}\\
&&\vec{s}=\frac{\vec{r}-\vec{r}_e}{|\vec{r}-\vec{r}_e|},\quad
\vec{s}'=-\frac{\vec{r}'-\vec{r}_e}{|\vec{r}'-\vec{r}_e|}.
\end{eqnarray}

%

\section{Phonons and Raman scattering}\label{sec:Phonons}

\begin{figure}
\includegraphics[width=8cm]{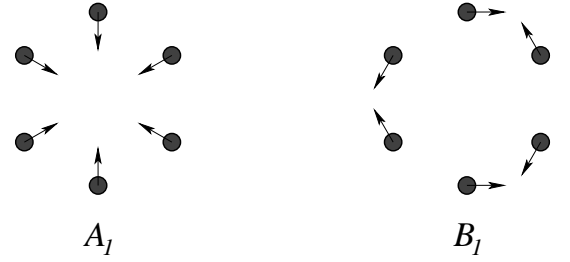}
\caption{\label{fig:phonons} Phonon modes responsible for the Raman
$D$~peak.}
\end{figure}

We restrict our attention to scalar phonons with wave vectors close to
$K$ and $K'$ points -- those responsible for the Raman $D$ peak. The
two real linear combinations of the modes at $K$ and $K'$ points
transform according to $A_1$ and $B_1$ representations of~$C_{6v}$ and
are shown in Fig.~\ref{fig:phonons}. We take the magnitude of the
carbon atom displacement as the normal coordinate for each mode,
denoted by $u_a$ and~$u_b$, respectively. Upon quantization of the
phonon field, the displacement operators $\hat{u}_a,\hat{u}_b$ and
the lattice hamiltonian $\hat{H}_{\mathrm{ph}}$ are expressed in
terms of the phonon creation and annihilation operators
$\hat{b}^\dagger_{\vec{q}\phonon},\hat{b}_{\vec{q}\phonon}$,
$\phonon=a,b$, as
\begin{subequations}\begin{eqnarray}
&&\hat{u}_\phonon(\vec{r})=L_xL_y\int\frac{d^2\vec{q}}{(2\pi)^2}
\frac{\hat{b}_{\vec{q}\phonon}e^{i\vec{q}\vec{r}}
+\hat{b}_{\vec{q}\phonon}^\dagger e^{-i\vec{q}\vec{r}}}
{\sqrt{2NM\omph}},\\
&&\hat{H}_{\mathrm{ph}}=L_xL_y\int\frac{d^2\vec{q}}{(2\pi)^2}
\sum_{\phonon=a,b}\omph
\left(\hat{b}_{\vec{q}\phonon}^\dagger\hat{b}_{\vec{q}\phonon}
+\frac{1}2\right).
\label{Hph=}
\end{eqnarray}\end{subequations}
The crystal is assumed to have the area $L_xL_y$, and to contain
$N$~carbon atoms of mass~$M$.
The area per carbon atom is $L_xL_y/N=\sqrt{27}a^2/4$.

The phonon frequency $\omph\approx{1350}\:\mbox{cm}^{-1}$, standing in
Eq.~(\ref{Hph=}), is assumed to be independent of the phonon momentum.
To check the validity of this assumption one should compare the
corresponding energy scale (the spread of the phonon momenta $\Delta{q}$
multiplied by the phonon group velocity~$v_\mathrm{ph}$) with the
electronic energy uncertainty. The latter is given by $\omph$ itself.
Recalling that phonon emission corresponds to the backscattering of
the electron (hole), $\Delta{q}$ is given by the uncertainty of the
electronic momentum, $\Delta{q}\sim\omph/v$. Since
$v_\mathrm{ph}/v\approx{7}\cdot{10}^{-3}\ll{1}$, the phonon dispersion
can be safely neglected.

If we neglect the phonon dispersion, the normal modes and the phonon
hamiltonian can be rewritten in the coordinate representation by
introducing the creation and annihilation operators for a phonon
in a given point of the sample:
\begin{subequations}\begin{eqnarray}
&&\hat\Phi_\phonon(\vec{r})=
\sum_\vec{q}\frac{\hat{b}_{\vec{q},\phonon}e^{i\vec{q}\vec{r}}}{\sqrt{L_xL_y}},\\
&&\hat{u}_\phonon(\vec{r})= \sqrt{\frac{L_xL_y}{2NM\omph}}
\left[\hat\Phi_\phonon(\vec{r})+\hat\Phi_\phonon^\dagger(\vec{r})\right],\\
&&\hat{H}_{\mathrm{ph}}= \sum_\phonon\omph\int{d}^2\vec{r}
\left[\hat\Phi_\phonon(\vec{r})\hat\Phi_\phonon^\dagger(\vec{r})
+\frac{N}{2L_xL_y}\right].
\end{eqnarray}\end{subequations}
Then it is convenient to define the phonon Green's function as
the time-ordered average of the $\Phi$-operators,
\begin{eqnarray}
&&D_{\phonon}^{(+)}(\vec{r},\omega)=-i\int\langle\mathrm{T}
\hat\Phi_\phonon(\vec{r},t)\,\hat\Phi^\dagger_{\phonon}(\vec{0},0)\rangle
e^{i\omega{t}}d^2\vec{r}\,dt=\nonumber\\
&&\qquad=\frac{\delta(\vec{r})} {\omega-\omph+io}.
\end{eqnarray}
By symmetry, in the electron-phonon interaction hamiltonian the
normal mode displacements $u_\phonon$ couple to the corresponding
valley-off-diagonal $4\times{4}$ matrices from
Table~\ref{tab:matrices}:\cite{megapaper}
\begin{equation}\label{Heph=}
\hat{H}_\mathrm{int}=F_K\int{d}^2\vec{r}\,\hat\psi^\dagger(\vec{r})
\left[\hat{u}_a(\vec{r})\Lambda_x\Sigma_z
+\hat{u}_b(\vec{r})\Lambda_y\Sigma_z\right] \hat\psi(\vec{r}).
\end{equation}
Here $F_K$ is the coupling constant having the dimensionality of
a force.
It is more convenient to use the dimensionless coupling constant
\begin{equation}\label{lambdaphonon=}
\lambda_K=
\frac{F_K^2}{M\omph v^2}\frac{\sqrt{27}a^2}4.
\end{equation}
The value of~$\lambda_K$ was discussed in Sec.~\ref{sec:qualpos}.

The hamiltonian describing interaction of electrons with light is
obtained from the Dirac hamiltonian~(\ref{Hel=}) by replacement
$\vec\nabla\to\vec\nabla-i(e/c)\hat{\vec{A}}$, where the vector
potential~$\hat{\vec{A}}$ is expressed in terms of creation and
annihilation operators
$\hat{a}^\dagger_{\vec{Q},\ell},\hat{a}_{\vec{Q},\ell}$ of
three-dimensional photons in the quantization volume $V=L_xL_yL_z$,
labeled by the three-dimensional wave vector~$\vec{Q}$ and two
transverse polarizations
$\ell=1,2$ with unit vectors $\vec{e}_{\vec{Q},\ell}$:
\begin{equation}
\hat{\vec{A}}(\vec{r})= \sum_{\vec{Q},\ell}\sqrt{\frac{2\pi{c}}{VQ}}
\left(\vec{e}_{\vec{Q},\ell}\hat{a}_{\vec{Q},\ell}e^{i\vec{Q}\vec{r}}
+\mathrm{h.c.} \right).\label{Afree=}
\end{equation}

The derivation of the formal expression for the Raman scattering probability
is fully analogous to that given in Ref.~\onlinecite{megapaper}. The only
difference is that the calculation is done in the coordinate representation.
As a result, we obtain the following expression for the probability for an
incident photon with wave vector~$\vec{Q}_{in}$ and polarization~$\vec{e}_{in}$
to be scattered with emission of a single phonon within an elementary
area~$d^2\vec{R}$ around a given point~$\vec{R}$:
\begin{subequations}
\begin{eqnarray}
\frac{dI_D}{d^2\vec{R}}&=&\frac{1}{cL_xL_y}\sum_{\vec{e}_{out}}
\int\frac{d^3\vec{Q}_{out}}{(2\pi)^3}\,
2\pi\delta(c|\vec{Q}_{out}|-\omega_{out})\nonumber\\
&&{}\times\sum_{\phonon=a,b}|2\mathcal{M}_\phonon|^2,\label{dId2Rgeneral=}\\
\mathcal{M}_\phonon&=&\sqrt{\frac{\lambda_K}2}
\frac{2\pi{e}^2v^3}{\sqrt{\omega_{in}\omega_{out}}}
\int\frac{d\ep}{2\pi}\,{d}^2\vec{r}_{in}\,{d}^2\vec{r}_{out}\nonumber\\
&&{}\times e^{i\vec{Q}_{in}\vec{r}_{in}-i\vec{Q}_{out}\vec{r}_{out}}
\Tr\left\{\mathcal{D}_\phonon+\bar{\mathcal{D}}_\phonon\right\},
\label{M1ph=}\\
\mathcal{D}_{a,b}&=&G(\vec{r}_{out},\vec{r}_{in};\ep)
(\vec{e}_{in}\cdot\vec\Sigma)\,
G(\vec{r}_{in},\vec{R};\ep-\omega_{in})\,\nonumber\\
&&{}\times\Lambda_{x,y}\Sigma_z\,
G(\vec{R},\vec{r}_{out};\ep-\omega_{out})(\vec{e}_{out}^*\cdot\vec\Sigma),
\label{D1tr=}\\
\bar{\mathcal{D}}_{a,b}&=&G(\vec{r}_{in},\vec{r}_{out};\ep)\,
(\vec{e}_{out}^*\cdot\vec\Sigma)\,
G(\vec{r}_{out},\vec{R};\ep+\omega_{out})\,\nonumber\\
&&{}\times\Lambda_{x,y}\Sigma_z\,
G(\vec{R},\vec{r}_{in};\ep+\omega_{in})(\vec{e}_{in}\cdot\vec\Sigma).
\label{D2tr=}
\end{eqnarray}\end{subequations}
Here $G(\vec{r},\vec{r}';\ep)$ is the electronic Green's function
corresponding to the full single-particle part of the hamiltonian
(i.~e., including not only the Dirac term, but the edge as well).
It can be represented in terms of the exact single-electron
eigenfunctions $\psi_s(\vec{r})$ and energies $\ep_s$ as a sum
over the eigenstates~$s$:
\begin{equation}
G(\vec{r},\vec{r}';\ep)=\sum_s\frac{\psi_s(\vec{r})\,\psi^\dagger_s(\vec{r})}
{\ep-\ep_s+i\gamma_\ep\sign\ep}.
\end{equation}
Using this representation, integrating over the energy and the
coordinates one obtains Eq.~(\ref{Ramanmatrixelement=}).
The summation in Eq.~(\ref{dId2Rgeneral=}) is performed over the
wave vectors $\vec{Q}_{out}$ and the polarizations $\vec{e}_{out}$
of the scatered photon. When integrated over the area of the crystal,
Eq.~(\ref{dId2Rgeneral=}) gives the absolute dimensionless probability
of the one-phonon Raman scattering for a single incident photon.

The matrix element $\mathcal{M}_\phonon$ can always be represented in the form
\begin{equation}
\mathcal{M}_\phonon=\mathcal{M}_\phonon^x(e_{out}^x)^*
+\mathcal{M}_\phonon^y(e_{out}^y)^*.
\end{equation}
If one collects all the light scattered in the full solid angle $4\pi$,
without analyzing the polarization, the integration over the angles of
$\vec{Q}_{out}$ is straightforward. It gives
\begin{subequations}\begin{equation}
\frac{dI_D}{d^2\vec{R}}=\frac{8}{3\pi}\,
\frac{\omega_{out}^2}{c^4L_xL_y}\, \sum_{\phonon=a,b}
\left(|\mathcal{M}_\phonon^x|^2+|\mathcal{M}_\phonon^y|^2\right).
\label{dId2total=}
\end{equation}
The dependence on the polarization of the scattered light is obtained
most easily when the light is collected in a small solid
angle~$o_{out}\ll{4\pi}$ around the normal (the case of an arbitrary
solid angle was considered in Ref.~\onlinecite{megapaper}). If the
analyzer is oriented at an angle $\varphi_{out}$ to the $x$~axis, the
polarization-dependent intensity is given by
\begin{equation}
\frac{dI_D}{d^2\vec{R}}=\frac{o_{out}}{\pi^2}
\frac{\omega_{out}^2}{c^4L_xL_y} \sum_{\phonon=a,b}
\left|\mathcal{M}_\phonon^x\cos\varphi_{out}+
\mathcal{M}_\phonon^y\sin\varphi_{out}\right|^2.
\label{dId2phiout=}
\end{equation}\end{subequations}

Eq.~(\ref{Afree=}) corresponds to the free-space quantization of the
electromagnetic field whose normal modes are plane waves. In the case
of a spatially resolved experiment as in Ref.~\onlinecite{Novotny},
Eqs. (\ref{dId2Rgeneral=}),~(\ref{M1ph=}) in order to account for the
spatial profile of the electric field, induced by the focusing lens.
Namely, the electric field, corresponding to a single photon with a
wave vector~$\vec{Q}_{in}$, incident from vacuum, should be replaced
by the field~$\mathcal{E}_{in}(\vec{r}_{in})$ of the focused laser beam,
\begin{equation}
i\sqrt{\frac{2\pi\omega_{in}}{L_xL_yL_z}}\,\vec{e}_{in}e^{i\vec{Q}_{in}\vec{r}_{in}}
\to\vec{e}_{in}\mathcal{E}_{in}(\vec{r}_{in}).
\end{equation}
As long as the distance between the lens and the sample is much larger
than the light wavelegth, the summation over the continuum of the final
states of the scattered photon can still be performed using the vacuum
mode structure.
Finally, dividing the resulting probability by the photon attempt
period $L_z/c$, we obtain the number of photons emitted per unit time,
$dI_D/dt$ (which is more appropriate when the incident light is
characterized by its electric field strength). As a result,
Eqs.~(\ref{dId2Rgeneral=}),~(\ref{M1ph=}) are modified as follows:
\begin{subequations}
\begin{eqnarray}
\frac{dI_D}{d^2\vec{R}\,dt}&=&\frac{1}{2\pi\omega_{in}}\sum_{\vec{e}_{out}}
\int\frac{d^3\vec{Q}_{out}}{(2\pi)^3}\,
2\pi\delta(c|\vec{Q}_{out}|-\omega_{out})\nonumber\\
&&{}\times\sum_{\phonon=a,b}|2\mathcal{M}_\phonon|^2,\label{dId2Rdtgeneral=}\\
\mathcal{M}_\phonon&=&\sqrt{\frac{\lambda_K}2}
\frac{2\pi{e}^2v^3}{\sqrt{\omega_{in}\omega_{out}}}
\int\frac{d\ep}{2\pi}\,{d}^2\vec{r}_{in}\,{d}^2\vec{r}_{out}\nonumber\\
&&{}\times \mathcal{E}_{in}(\vec{r}_{in})\,e^{-i\vec{Q}_{out}\vec{r}_{out}}
\Tr\left\{\mathcal{D}_\phonon+\bar{\mathcal{D}}_\phonon\right\}.
\label{M1phE=}
\end{eqnarray}\end{subequations}

\section{Raman scattering on a translationally invariant edge}\label{sec:Regular}

\subsection{General considerations}\label{sec:reggeneral}

For simplicity we consider an armchair edge characterized by
$\vec{n}=\vec{e}_x$,  $M=-\Sigma_y\Lambda_y$, as discussed in the previous
section. Due to the translational invariance in the $y$~direction, it is
sufficient to calculate the probability of phonon emission at the point
$\vec{R}=X\vec{e}_x$. As we will see below, the main contribution to the
signal comes from $X\gg\lep=2v/\omega_{in}$. In this regime the motion
of the photoexcited electron-hole pair can be described quasiclassically,
and the asymptotic large-distance expansion for the Green's functions,
Eq.~(\ref{GreenfLarger=}), can be used.
Namely, the electron and the hole can be viewed as wave packets of the size
$\sim\sqrt{\lep{X}}$, propagating across the crystal along classical
trajectories. Initially, they are created in the same region of space
of the size $\sim\sqrt{\lep{X}}$ with opposite momenta and opposite
velocities. As they undergo scattering processes (emission of a phonon or
reflection from the edge), they change the directions of their momenta.
In order to recombine radiatively and contribute to Raman signal,
the electron and the hole should meet again within a spatial region of the
size $\sim\sqrt{\lep{X}}$. Clearly, these conditions can be fulfilled if
all momenta are almost perpendicular to the edge. Small deviations by an
angle $\sim\sqrt{\lep/X}$ are allowed by quantum uncertainty.
These considerations are illustrated by Fig.~\ref{fig:traject}.

Emission of one of the phonons shown in Fig.~\ref{fig:phonons} corresponds
to intervalley scattering of the photoexcited electron or the hole, as
represented formally by one of the valley-off-diagonal matrices $\Lambda_x$
or $\Lambda_y$ in the matrix element, see Eqs.~(\ref{M1ph=})--(\ref{D2tr=}).
For the process to be allowed, another act of intervalley scattering is
required, so one of the three electronic Green's functions in
Eqs.~(\ref{D1tr=}), (\ref{D2tr=}) must contain another $\Lambda_x$ or
$\Lambda_y$ matrix from the decomposition~(\ref{imagesource=}). Otherwise,
the trace in Eq.~(\ref{M1ph=}) vanishes. Thus, for an armchair edge with
$M=-\Lambda_y\Sigma_y$ only the $B_1$ phonon can be emitted, so
$\mathcal{M}_a=0$, $\mathcal{M}_b=\mathcal{M}$.
Trajectories, corresponding to
decomposition of each of the three Green's functions in Eq.~(\ref{D1tr=})
are shown in Fig.~\ref{fig:traject} (a), (b), (c), respectively. According
to the quasiclassical picture, the electron and the hole have to travel the
same distance between creation and annihilation, as their velocities are
equal. Then the process in Fig.~\ref{fig:traject}~(a) has more phase space
satisfying this restriction, and this process gives the main contribution to
the Raman matrix element. This will be also shown explicitly below.

The general polarization structure of the matrix element, compatible with
the reflection symmetry $y\to{-}y$ possessed by the armchair edge, is
\begin{eqnarray}
\mathcal{M}&=&\mathcal{M}_\|e_{in}^y(e_{out}^y)^* +\mathcal{M}_\perp
e_{in}^x(e_{out}^x)^*\nonumber\\
&=&\mathcal{M}_\|\sin\varphi_{in}\sin\varphi_{out}
+\mathcal{M}_\perp\cos\varphi_{in}\cos\varphi_{out}\label{Mparperp=}
\end{eqnarray}
(we introduced $\varphi_{in},\varphi_{out}$, the polar angles of the
polarization vectors). Since the
interband transition dipole moment is perpendicular to the electronic
momentum, and for a regular edge the momentum is (almost)
perpendincular to the edge, $\mathcal{M}_\perp$ is entirely due to
quantum diffraction. Thus, $\mathcal{M}_\perp$ is smaller than
$\mathcal{M}_\|$ by the parameter $\lep/X\ll{1}$ (it is this parameter
that governs the quantum diffraction, as discussed above).
Nevertheless, in a polarization-resolved experiment the two intensities
can be measured independently, so below we calculate both
$\mathcal{M}_\|$ and $\mathcal{M}_\perp$, each to the leading order in
$\lep/X$.


\subsection{Spatially integrated intensity and polarization dependence}
\label{sec:integrated}

In this subsection the electric field of the excitation wave is assumed to be
spatially homogeneous. As displacements along the edge are expected to be
parametrically smaller than those away from the edge, $|y|\sim\sqrt{x\lep}$,
we use the paraxial approximation, $|\vec{r}|\approx|x|+y^2/(2|x|)$.
The Green's function can be approximated as
\begin{eqnarray}
G_0(\vec{r},\ep)&\approx&-\sqrt{\frac{i|\ep|}{2\pi{v}^3|x|}}\,
e^{(i|\ep|-\gamma)|x|/v+i|\ep|y^2/(2v|x|)}\nonumber\\
&&{}\times\left[\frac{\sign\ep+\Sigma_x\sign{x}}{2}\right.\nonumber\\
&&\;\;{}+\left.\frac{y\Sigma_y}{2|x|}
+\frac{\sign\ep-\Sigma_x\sign{x}}{8}
\left(\frac{y^2}{x^2}-\frac{iv}{|\ep{x}|}\right)\right],
\nonumber\\ && \label{G0parax=}
\end{eqnarray}
where the coefficient at each matrix is taken in the leading order.
Taking the first term in the square brackets in Eq.~(\ref{G0parax=}) and
evaluating the matrix traces, we obtain the following expression for
$\mathcal{M}_\|$, corresponding to the process in Fig.~\ref{fig:traject}~(a):
\begin{subequations}\begin{eqnarray}
\mathcal{M}_\|&=&
\sqrt{\frac{i\lambda_K}{\pi^3v}}\,\frac{e^2}{\omega_{in}}\nonumber\\
&&{}\times\int\limits_{-\infty}^\infty\frac{d\ep}v
\int\limits_0^Xdx_{in}dx_{out}
\int\limits_{-\infty}^\infty{d}y_{in}dy_{out}\,e^{i\Phi-2\gamma{X}/v}\nonumber\\
&&{}\times\sqrt{\frac{|(\omega_{in}-\ep)(\omega_{out}-\ep)\ep|}
{(X-x_{in})(X-x_{out})(x_{in}+x_{out})}},\label{Minitial=}\\
\Phi&=&\frac{|\ep|({x}_{in}+x_{out})}v
+\frac{|\ep|(y_{in}-y_{out})^2}{2v(x_{in}+x_{out})}
+\nonumber\\
&&{}+\frac{|\omega_{out}-\ep|(X-x_{out})}v
+\frac{|\omega_{out}-\ep|y_{out}^2}{2v(X-{x}_{out})}+\nonumber\\
&&{}+\frac{|\omega_{in}-\ep|(X-x_{in})}v+\frac{|\omega_{in}-\ep|y_{in}^2}{2v(X-{x}_{in})}.\label{Phi=}
\end{eqnarray}\end{subequations}
For the calculation of $\mathcal{M}_\perp$ we need the rest of the
terms in the square brackets in Eq.~(\ref{G0parax=}). As a result,
$\mathcal{M}_\perp$ is given
by an analogous integral, but with an extra factor in the integrand,
\[\begin{split}
&\left[\frac{1}4
\left(\frac{y_{out}-y_{in}}{x_{in}+x_{out}}+\frac{y_{in}}{X-x_{in}}\right)
\left(\frac{y_{in}-y_{out}}{x_{in}+x_{out}}+\frac{y_{out}}{X-x_{out}}\right)\right.\\
&+\left.\frac{iv}{4|\ep|(x_{in}+x_{out})}\right].
\end{split}\]
The details of integration are given in Appendix~\ref{app:integrals}.
First, we integrate over $y_{in}$ and $y_{out}$.
The subsequent integration over~$\ep$ fixes
$x_{in}/v+x_{out}/v\approx(X-x_{in})/v+(X-x_{out})/v$ [the difference
is allowed to be $\sim\sqrt{X/(v\omega_{in})}\ll{X}$], which has the
meaning of the time spent by the electron and the hole in traveling
from the creation point~$x_{in}$ to the annihilation point~$x_{out}$.
At the same time, $x$-integration fixes $\ep\approx\omega_{in}/2$ with
the precision $\sim\sqrt{\omega_{in}v/X}\ll\omega_{in}$. Performing all
the integrations, we obtain the matrix element,
\begin{subequations}\begin{eqnarray}
\mathcal{M}_\|&=&
e^2\sqrt{\frac{\pi\lambda_K}{i\omega_{in}X/v}}\,
\frac{\sin[\omph{X}/(2v)]}{\omph/(2v)}
\times\nonumber\\&&{}\times
{e}^{i(\omega_{in}+\omega_{out})X/(2v)-2\gamma{X}/v},
\label{Mfinal=}\\
\mathcal{M}_\perp&=&\frac{i\mathcal{M}_\|}{\omega_{in}X/v}.
\label{Mperpfinal=}
\end{eqnarray}\end{subequations}
According to Eq.~(\ref{dId2total=}), the integrated intensity into the
full solid angle $4\pi$, summed over two polarizations of the emitted
photon, is given by
\begin{subequations}\begin{eqnarray}
\frac{d^2I_D}{d^2\vec{R}}&=&
\frac{8\lambda_K}{3}\,\frac{(e^2/c)^2}{{L}_xL_y}\,
\frac{\omega_{in}^2}{c^2}\,\frac{\sin^2[\omph{X}/(2v)]}{[\omph/(2v)]^2}\,
\frac{e^{-4\gamma{X}/v}}{\omega_{in}X/v}\times\nonumber\\&&{}\times
\left[\sin^2\varphi_{in}+{(\omega_{in}X/v)^2}\,\cos^2\varphi_{in}\right],
\label{IDregulara=}\\
I_D&=&\frac{8\lambda_K}{3}\left(\frac{e^2}c\right)^2\frac{v^2}{c^2}\,
\frac{v}{\omega_{in}L_x}\,\frac{\omega_{in}^2}{\omph^2}
\times\nonumber\\&&{}\times
\left[\sin^2\varphi_{in}\ln\frac{\omph^2+(4\gamma)^2}{(4\gamma)^2}\right.+\nonumber\\
&&{}\qquad+\left.\cos^2\varphi_{in}\,\frac{\omph^2}{\omega_{in}^2}
\ln\frac{\omega_{in}}\omph\right].\label{IDregularb=}
\end{eqnarray}
The second term in the square brackets is written with logarithmic
precision, since the short-distance cutoff $\sim{v}/\omega_{in}$ is
known only up to a factor of the order of 1. If we use
Eq.~(\ref{dId2phiout=}) for the intensity in a solid angle $o_{out}$ in
the presence of an analyzer, we obtain
\begin{eqnarray}
I_D&=&4\lambda_K\,\frac{o_{out}}{4\pi}
\left(\frac{e^2}c\right)^2\frac{v^2}{c^2}\,
\frac{v}{\omega_{in}L_x}\,\frac{\omega_{in}^2}{\omph^2}
\times\nonumber\\&&{}\times
\left[\sin^2\varphi_
{in}\sin^2\varphi_{out}\ln\frac{\omph^2+(4\gamma)^2}{(4\gamma)^2}\right.+\nonumber\\
&&{}\qquad+\left.\cos^2\varphi_{in}\cos^2\varphi_{out}\,\frac{\omph^2}{\omega_{in}^2}
\ln\frac{\omega_{in}}\omph\right].
\label{IDregularc=}
\end{eqnarray}
\end{subequations}

Let us estimate the contribution to the matrix element~$\mathcal{M}'$
corresponding to Fig.~\ref{fig:traject}~(b), i.~e., when
decomposition~(\ref{imagesource=}) is applied to
$G(\vec{R},\vec{r}_{out};\ep-\omega_{out})$.
Eq.~(\ref{Mparperp=}) remains valid, as it is based on symmetry only.
The expression for~$\mathcal{M}_\|'$ looks analogous to
Eqs.~(\ref{Minitial=}), (\ref{Phi=}):
\begin{subequations}\begin{eqnarray}
\mathcal{M}_\|^\prime&=&\frac{e^2}{\omega_{in}}\,
\sqrt{\frac{i\lambda_K}{\pi^3v}}\int\limits_{-\infty}^\infty\frac{d\ep}v
\times\nonumber\\
&&{}\times
\int\limits_0^Xdx_{in}\int\limits_0^{x_{in}}dx_{out}
\int\limits_{-\infty}^\infty{d}y_{in}dy_{out}\,e^{i\Phi'-2\gamma{X}/v}\times\nonumber\\
&&{}\times\sqrt{\frac{|(\omega_{in}-\ep)(\omega_{out}-\ep)\ep|}
{(X-x_{in})(X+x_{out})(x_{in}-x_{out})}},\label{Mprimea=}\\
\Phi'&=&\frac{|\ep|({x}_{in}-x_{out})}v
+\frac{|\ep|(y_{in}-y_{out})^2}{2v(x_{in}-x_{out})}
+\nonumber\\
&&{}+\frac{|\omega_{out}-\ep|(X+x_{out})}v
+\frac{|\omega_{out}-\ep|y_{out}^2}{2v(X+{x}_{out})}+\nonumber\\
&&{}+\frac{|\omega_{in}-\ep|(X-x_{in})}v+\frac{|\omega_{in}-\ep|y_{in}^2}{2v(X-{x}_{in})}.
\label{Mprimeb=}
\end{eqnarray}\end{subequations}
However, here integration over~$\ep$ fixes
$x_{in}-x_{out}\approx{x}_{out}+X+(X-x_{in})$, which is compatible with
the limits of the spatial integration only when
$x_{out}\sim{v}/\omega_{in}$, $X-x_{in}\sim{v}/\omega_{in}$, as shown
in Fig~\ref{fig:traject}~(b). This restriction results in suppression
$\mathcal{M}_\|'/\mathcal{M}_\|\sim\omph/\omega_{in}\ll{1}$. 

If $x_{out},|X-x_{in}|\sim{v}/\omega_{in}$, the asymptotic form,
Eq.~(\ref{G0parax=}), cannot be used for
$G(\vec{r}_{in},\vec{R};\omega_{in}-\ep)$
[thus, Eqs.~(\ref{Mprimea=}),~(\ref{Mprimeb=}) represent only an
estimate of $\mathcal{M}_\|'$ by the order of magnitude], but it can
be used for $G(\vec{R},\vec{r}_{out};\omega_{out}-\ep)$,
and $G(\vec{r}_{out},\vec{r}_{in};\ep)$. This fact results in an
additional smallness for the matrix element $\mathcal{M}_\perp'$:
assuming the typical value $X\sim{v}/\omph$, we can write
$|\mathcal{M}_\perp'|\sim|\mathcal{M}_\|'|(\omph/\omega_{in})\sim%
|\mathcal{M}_\perp|(\omph/\omega_{in})$.
Thus, $\mathcal{M}_\perp'$ produces only a small correction to the
$\cos^2\varphi_{in}$ term in Eqs.~(\ref{IDregulara=}),~(\ref{IDregularb=}),
and to the $\cos^2\varphi_{in}\cos^2\varphi_{out}$ term in
Eq.~(\ref{IDregularc=}).

Finally, the intensity in Eq.~(\ref{IDregularc=}) has an
interference contribution
$\propto\Re(\mathcal{M}_\|^*\mathcal{M}_\perp')$, which
produces a term
$\propto\sin\varphi_{in}\cos\varphi_{in}\sin\varphi_{out}\cos\varphi_{out}$.
Note that the interference term $\Re(\mathcal{M}_\|^*\mathcal{M}_\perp)$
is absent because of the factor~$i$ in Eq.~(\ref{Mperpfinal=}). We have not
been able to calculate $\mathcal{M}_\perp'$ explicitly or to establish a
phase relationsip between $\mathcal{M}_\|$ and $\mathcal{M}_\perp'$ in
any other way.
However, it is hard to imagine that the inteference of two
strongly oscillating amplitudes, corresponding to two different
processes, would survive the integration over~$X$.

\subsection{Excitation position dependence}
\label{sec:regSpatial}

This subsection aims at describing a spatially resolved experiment like
that of Ref.~\onlinecite{Novotny} and clarifying the role of different
length scales in the dependence of the Raman intensity on the position
of the excitation spot.
Consequently, we use Eqs.~(\ref{dId2Rdtgeneral=}),~(\ref{M1phE=}), and
repeat the calculation of Sec.~\ref{sec:integrated} with an arbitrary
dependence of $\mathcal{E}_{in}(\vec{r})$, smooth on the length scale
$v/\omega_{in}$ (we assume detection without analyzer and sum over the
polarizations of the scattered photon). The result is
\begin{subequations}\begin{eqnarray}
&&\frac{dI_D}{dt}=
\frac{4\lambda_K}{3\pi}\left(\frac{e^2}c\right)^2\frac{v}{c}\,\sin^2\varphi_{in}
\int\limits_{-\infty}^\infty{d}y\,
\mathcal{I}\!\left(\frac{v}{\omph},\frac{v}{2\gamma}\right),
\label{dIDdtreg=}\\
&&\mathcal{I}(\ellph,\ell_\gamma)=\int\limits_0^\infty
\mathcal{E}_{in}(x,y)\,\mathcal{E}_{in}^*(x',y)\,\mathcal{K}(x,x')\,
dx\,dx',\label{Inonlocal=}\\
&&\mathcal{K}(x,x')=-e^{-i(x-x')/\ellph}\Ei(-2\max\{x,x'\}/\ell_\gamma),
\label{spatialkernel=}
\end{eqnarray}
where the exponential integral $\Ei(z)$ is defined as
\begin{equation}
-\Ei(-z)=\int\limits_z^\infty\frac{e^{-t}\,dt}t.
\end{equation}\end{subequations}
Let us assume the excitation intensity to have the form
$|\mathcal{E}_{in}(x)|^2=w(x-x_0)$, where $w(x)$ is some smooth
bell-shaped function centered at $x=0$, and $x_0$~is the position of
the focus, which serves as the experimental control parameter.
In the following we also assume that
the phase of $\mathcal{E}_{in}(x)$ does not depend on~$x$, then
$\mathcal{E}_{in}(x)$ can be taken real without loss of generality.
The integral in Eq.~(\ref{Inonlocal=}) is
determined by three length scales. The width~$L$ of the excitation
profile $w(x)$ is assumed to be much longer than $\ellph=v/\omph$
and the electron inelastic scattering length $\ell_\gamma=v/(2\gamma)$:
$\ellph,\ell_\gamma\ll{L}$. In all above expressions of this section
no assumption was made about the relative magnitude of
$\ellph$~and~$\ell_\gamma$. However, it is reasonable to assume
$\ellph\lesssim\ell_\gamma$;
also, the final expressions become more compact in this limit.

In the zeroth approximation in $1/L$ one can assume
$w(x)=\mathrm{const}$ in the effective region of integration in
Eq.~(\ref{Inonlocal=}), i.~e. replace the kernel by a
$\delta$-function. This gives the result of Sec.~\ref{sec:integrated},
\begin{subequations}\begin{eqnarray}
&&\mathcal{I}_{x_0}= l_0^2w(-x_0)+O(\ell/L),\\
&&l_0^2=\int\limits_0^\infty\mathcal{K}(x,x')\,dx\,dx'
=\ellph^2\ln\frac{\ell_\gamma^2+4\ellph^2}{4\ellph^2}.
\end{eqnarray}\end{subequations}
The length scale $l_0$, appearing here, may be viewed as the effective
range of integration in Eq.~(\ref{Inonlocal=}), which determines the
magnitude of the signal. As we see, this length is mainly determined by
$\ellph$ and is only logarithmically sensitive to the electron
inelastic scattering.

What is detected experimentally in Ref.~\onlinecite{Novotny} is the
difference in the profiles of $w(-x_0)$ and $\mathcal{I}_{x_0}$,
appearing because of the non-locality of the kernel.
This difference appears in the second order of the expansion of
Eq.~(\ref{Inonlocal=}) in the spatial derivatives of
$\mathcal{E}_{in}(x)$ (i.~e, in the order $1/L^2$):
\begin{equation}
\mathcal{I}_{x_0}=l_0^2w(l_1-x_0)+\frac{l_0^2l_2^2}2\,w''(l_1-x_0)
+O(\ell^3/L^3).\label{expandprofile=}
\end{equation}
Here the length $l_1$ [the ``center of mass'' of the kernel in
Eq.~(\ref{spatialkernel=})] is given by
\begin{eqnarray}
l_1&=&\Re\int\limits_0^\infty{x}\,\mathcal{K}(x,x')\,\frac{dx\,dx'}{l_0^2}
\nonumber\\
&=&\frac{\ell_\gamma^3/8}{\ell_\gamma^2/4+\ellph^2}
\left(\ln\frac{\ell_\gamma^2/4+\ellph^2}{\ellph^2}\right)^{-1}.
\end{eqnarray}
It describes the overall shift of the profile, which may be difficult
to detect experimentally, unless the precise location of the edge is known.
The length $l_2$, appearing in Eq.~(\ref{expandprofile=}), determines the
broadening of the signal profile $\mathcal{I}_{x_0}$ with respect to
the excitation profile $w(-x_0)$, proportional to $w''(x)$ (the second
derivative). In the limit $\ell_\gamma\gg\ellph$ it is given by
(see Appendix~\ref{app:kernel} for the full expression and other details):
\begin{equation}\label{l2=}
l_2^2
=\ell_\gamma^2
\frac{2\ln(\ell_\gamma/\ellph)-1}{16\ln^2(\ell_\gamma/\ellph)}
+O(\ellph^2).
\end{equation}
Note that this length is indeed determined by the electronic
inelastic length (up to logarithmic corrections), in qualitative
agreement with the assumption of Ref.~\onlinecite{Novotny}.

Instead of Eq.~(\ref{spatialkernel=}), Can\c{c}ado, Beams and
Novotny\cite{Novotny} fitted the experimental profile $\mathcal{I}_{x_0}$
using the following expression:
\begin{equation}\label{wrongkernel=}
\mathcal{I}_{x_0}\propto\left|\int\limits_{x_e}^\infty
e^{-(x-x_e)/x_D}\mathcal{E}_{in}(x-x_0)\,dx\right|^2,
\end{equation}
where the excitation profile was independently determined to be
gaussian: $|\mathcal{E}_{in}(x)|^2\propto{e}^{-x^2/L^2}$. The effective
position of the edge~$x_e$, the width~$x_D$, as well as the overall
proportionality coefficient were used as fitting parameters, and the
value $x_D=20$~nm was obtained.  Expanding Eq.~(\ref{wrongkernel=}) to
the order $1/L^2$ and comparing it to Eq.~(\ref{expandgaussian=}), we
obtain
\begin{equation}\begin{split}
&Ae^{-\frac{(x_0-x_e-x_D)^2}{L^2}}
\left[1+\frac{x_D^2}{L^2}\left(\frac{(x_0-x_e-x_D)^2}{L^2}-1\right)\right]=\\
&=e^{-\frac{(x_0-l_1)^2}{L^2}}
\left[1+\frac{l_2^2}{L^2}\left(\frac{2(x_0-l_1)^2}{L^2}-1\right)\right]
+O(L^{-3}).
\end{split}\end{equation}
This equation is satisfied for all~$x_0$ provided that $x_e+x_D=l_1$,
$A=1+l_2^2/L^2$, $x_D^2=2l_2^2$. Thus, in spite of the fact that in
Ref.~\onlinecite{Novotny} a wrong kernel was used, we still can take
the experimentally found~$l_2$ and use it with the correct kernel.
Namely, the experimentally
measured $x_D=20$~nm yields $l_2=14$~nm. Using Eq.~(\ref{l2=}) and
taking $\ellph=4$~nm, we obtain $\ell_\gamma=66$~nm, which gives
$2\gamma=11$~meV. As discussed in Sec.~\ref{sec:qualpos} and in
Ref.~\onlinecite{Casiraghi2009}, this value is significantly smaller
than an estimate obtained using other sources of information.

\section{Raman scattering on an atomically rough edge}\label{sec:Rough}

In this section we calculate the Raman intensity for an edge rough
on atomic scale, and described by the model of Sec.~\ref{sec:irregular}.
The general arguments of Sec.\ref{sec:reggeneral} mostly remain valid,
except for that of the symmetry $y\to-y$, not possessed by any given
realization of the disorder. This symmetry is restored upon averaging
of the intensity over the realizations of disorder, but the
matrix element~$\mathcal{M}_\mu$ must be taken in the general form.

\subsection{Spatially integrated intensity and polarization dependence}
\label{sec:disordpolariz}

\begin{figure}
\includegraphics[width=8cm]{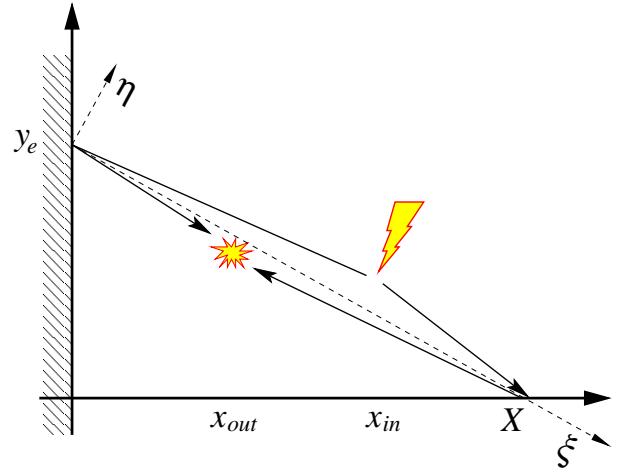}
\caption{\label{fig:trajectoblique}(Color on-line.) Electron
trajectories corresponding to Raman scattering on a disordered edge.
Notations are the same as in Fig.~\ref{fig:traject}.}
\end{figure}

Since the edge can scatter an electron by an arbitrary angle,
it is convenient to use the rotated coordinates $(\xi,\eta)$, as shown
in Fig.~\ref{fig:trajectoblique}. Taking the first Green's function in
Eqs.~(\ref{D1tr=}), (\ref{D2tr=}) as given by Eq.~(\ref{gfTmatrix=}),
and taking the free Green's functions in the paraxial approximation
(i.~e., assuming $|\eta|\ll\xi$), we arrive at the following expression
for the Raman matrix element:
\begin{subequations}\begin{eqnarray}
\mathcal{M}_\phonon&=&
\sqrt{\frac{\lambda_K}{2}}\,\frac{e^2v}{4\pi^2\omega_{in}}\times\nonumber\\
&&{}\times
\int\limits_{-\infty}^\infty{dy}_e\,e^{-2(\gamma/v)\sqrt{X^2+y_e^2}}
\Tr_{2\times{2}}\{\Lambda_\phonon{S}_\Lambda(y_e)\}\times\nonumber\\
&&{}\times
\frac{X(Xe_{in}^y+y_ee_{in}^x)(Xe_{out}^y+y_ee_{out}^x)^*}{(X^2+y_e^2)^{3/2}}\times\nonumber\\
&&{}\times\int\limits_{-\infty}^\infty\frac{d\ep}v
\int\limits_0^{\sqrt{X^2+y_e^2}}d\xi_{in}d\xi_{out}
\int\limits_{-\infty}^\infty{d}\eta_{in}d\eta_{out}\times\nonumber\\
&&{}\times \sqrt{\frac{|\omega_{in}-\ep||\omega_{out}-\ep||\ep|^2}
{v^4\xi_{in}\xi_{out}\xi_{in}'\xi_{out}'}}\,
e^{i\Phi_{in}+i\Phi_{out}},\label{Mirrinitial=}\\
\Phi_a&=&
\frac{|\ep|}v\left(\xi_a+\frac{\eta_a^2}{2\xi_a}\right)
+\frac{|\omega_a-\ep|}{v}\left(\xi_a'
+\frac{\eta_a^2}{2\xi_a'}\right),\\
\xi_a'&=&\sqrt{X^2+y_a^2}-\xi_a,
\end{eqnarray}\end{subequations}
where $a$ is either ``$in$'' or ``$out$''. Analogously to the case of
the regular edge, we first integrate over $\eta_{in},\eta_{out}$, and
subsequent integration over $\ep$ and $\xi_{in,out}$ fixes
$\xi_{in}+\xi_{out}\approx\sqrt{X^2+y_e^2}$, $\ep\approx\omega_{in}/2$.
As a result, we obtain
\begin{eqnarray}
\mathcal{M}_\phonon&=& \frac{ie^2}{4}\sqrt{\frac{\lambda_K}{2}}
\int\limits_{-\infty}^\infty{dy}_e\,
\Tr_{2\times{2}}\{\Lambda_\phonon{S}_\Lambda(y_e)\}
\times\nonumber\\
&&{}\times
e^{(i\omega_{in}/2+i\omega_{out}/2-2\gamma)\sqrt{X^2+y_e^2}/v}
\times\nonumber\\
&&{}\times
\frac{X(Xe_{in}^y+y_ee_{in}^x)(Xe_{out}^y+y_ee_{out}^x)^*}{(X^2+y_e^2)^2}\times\nonumber\\
&&{}\times\frac{\sin[\omph\sqrt{X^2+y_e^2}/(2v)]}{\omph/(2v)}.
\end{eqnarray}
To calculate the intensity, we use
Eqs.~(\ref{SLambdaCOE=}),~(\ref{COEaverage=}) to average the square of
the matrix element, and sum over the two phonon modes. Angular
integration and summation over the two detector polarizations according
to Eq.~(\ref{dId2total=}) gives
\begin{subequations}\begin{eqnarray}
\frac{dI_D}{d^2\vec{R}}&=&
\frac{2\lambda_K}{9}\left(\frac{e^2}c\right)^2\frac{v^2}{c^2}\,
\frac{\omega_{in}/v}{L_xL_y}\int\limits_{-\infty}^\infty{d}y_e\,
e^{-4\gamma\sqrt{X^2+y_e^2}/v}\times\nonumber\\
&&{}\times\frac{\sin^2[\omph\sqrt{X^2+y_e^2}/(2v)]}{[\omph/(2v)]^2}
\times\nonumber\\&&{}\times
\frac{X^2(X^2\sin^2\varphi_{in}+y_e^2\cos^2\varphi_{in})}{(X^2+y_e^2)^3},\\
I_D&=&
\frac{\pi\lambda_K}{36}\left(\frac{e^2}c\right)^2\frac{v^2}{c^2}\,
\frac{\omega_{in}}{vL_x}\,\frac{v^2}{\omph^2}
\ln\frac{\omph^2+(4\gamma)^2}{(4\gamma)^2}\times\nonumber\\
&&{}\times\left(3\sin^2\varphi_{in}+\cos^2\varphi_{in}\right).
\label{IDdisordb=}
\end{eqnarray}
Eq.~(\ref{dId2phiout=}) for the intensity emitted into a solid angle
$o_{out}$ in the presence of an analyzer gives
\begin{eqnarray}
I_D&=&
\frac{\pi\lambda_K}{24}\frac{o_{out}}{4\pi}
\left(\frac{e^2}c\right)^2\frac{v^2}{c^2}\,
\frac{\omega_{in}}{vL_x}\,\frac{v^2}{\omph^2}
\ln\frac{\omph^2+(4\gamma)^2}{(4\gamma)^2}\times\nonumber\\
&&{}\times\left[\sin^2\varphi_{in}+\sin^2\varphi_{out}
+\frac{1}2\cos(2\varphi_{in}-2\varphi_{out})\right].\nonumber\\
\label{IDdisordc=}
\end{eqnarray}
\end{subequations}
The trigonometric expression in the square brackets can be identically
rewritten as
$\sin^2\varphi_{in}+\sin(2\varphi_{out}-\varphi_{in})\sin\varphi_{in}+1/2$,
so its absolute minimum is 1/4, reached at
$\varphi_{out}=-\varphi_{in}=\pm\pi/6$.

\subsection{Excitation position dependence}
\label{sec:disordspatial}

Here we follow the same logic as in Sec.~\ref{sec:regSpatial},
but instead of Eqs.~(\ref{dIDdtreg=})--(\ref{spatialkernel=}) we have
\begin{eqnarray}
\frac{dI_D}{dt}&=&\frac{\lambda_K}{9\pi}\left(\frac{e^2}c\right)^2\frac{v}c
\nonumber\\ &&{}\times
\int\limits_{-\infty}^\infty dY\,dX\,dy_e\,
e^{-4(\gamma/v)\sqrt{X^2+y_e^2}}\,
\nonumber\\ &&{}\times
\frac{X^2(X\sin\varphi_{in}+y_e\cos\varphi_{in})^2}{(X^2+y_e^2)^3}
\nonumber\\ &&{}\times
\int\limits_0^{\sqrt{X^2+y_e^2}}d\xi_{in}\,d\xi_{in}'\,
e^{-i(\omph/v)(\xi_{in}-\xi_{in}')}
\nonumber\\ &&{}\times
\mathcal{E}_{in}\!\left(\frac{\xi_{in}X}{\sqrt{X^2+y_e^2}},Y+y_e-\frac{\xi_{in}y_e}{\sqrt{X^2+y_e^2}}\right)
\nonumber\\ &&{}\times
\mathcal{E}_{in}^*\!\left(\frac{\xi_{in}'X}{\sqrt{X^2+y_e^2}},Y+y_e-\frac{\xi_{in}'y_e}{\sqrt{X^2+y_e^2}}\right),
\end{eqnarray}
where $(X,Y)$ is the point where the phonon is emitted.
As in Sec.~\ref{sec:regSpatial}, we expand in the spatial derivatives
of $\mathcal{E}_{in}(\vec{r})$, and obtain Eq.~(\ref{expandprofile=})
with $l_2$ given by (actually, the result depends slightly on the
polarization; here we choose unpolarized detection and excitation
polarization along the edge, $\varphi_{in}=\pi/2$):
\begin{equation}
l_2^2=\ell_\gamma^2
\frac{5\ln(\ell_\gamma/\ellph)-3\cdot{2}^{14}/(45\pi)^2}{48\ln^2(\ell_\gamma/\ellph)}
+O(\ellph^2).
\end{equation}
This expression reproduces the experimental value $l_2=14\:\mbox{nm}$
if $\ell_\gamma=120\:\mbox{nm}$ (again, $\ellph=4\:\mbox{nm}$ is taken).

\section{Raman scattering on a fragmented edge}
\label{sec:fragmented}

In this section we consider the Raman scattering on an edge
consisitng of armchair and zigzag segments whose typical
length $d_e$ significantly exceeds the electronic wavelength,
$d_e\gg\lep$, where $\lep=2v/\omega_{in}$. This is an
intermediate case between the two limiting cases considered
in the previous two sections.

Only armchair segments contribute to the Raman
process.\cite{Cancado2004} Moreover, contributions from different
segments add up incoherently. Thus, we first focus on the
contribution of a single armchair segment, placed at $x=0$,
$-d_e/2\leq{y}\leq{d}_e/2$ (as before, graphene is assumed to
occupy the half-space $x>0$).
The electronic Green's function
corresponding to the reflection from a single armchair segment can be
easily written from the Huygens-Fresnel principle, Eq.~(\ref{Huygens=})
if one approximates its value on the boundary by that for an infinite
perfect armchair edge:
\begin{eqnarray}
&&{G}(\vec{r},\vec{r}';\ep)=G_0(\vec{r}-\vec{r}',\ep)
-iv\int\limits_{-d_e/2}^{d_e/2}dy_e\nonumber\\
&&\qquad\qquad{}\times G_0(x,y-y_e;\ep)\,
\Sigma_x\,G_0(x',y_e-y';\ep)\,\Sigma_y\Lambda_y.\nonumber\\
\end{eqnarray}
This approximation ignores the change of the exact wave function
within the distance $\sim\lep$ from the ends of the segment, which gives
a small error if $d_e\gg\lep$. In fact, it is the standard approximation
for the study of diffraction in optics.\cite{BornWolf}

Using this Green's function, we obtain the following expression for the
matrix element corresponding to emission of a phonon in an arbitrary
point $(X,Y)$:
\begin{subequations}\begin{eqnarray}
&&\mathcal{M}=
-\sqrt{\frac{\lambda_K}{2}}\,\frac{e^2v}{\pi^2\omega_{in}}
\int\limits_{-\infty}^\infty\frac{d\ep}v
\int{d}^2\vec{r}_{in}\,d^2\vec{r}_{out}\int\limits_{-d_e/2}^{d_e/2}dy_e\nonumber\\
&&\qquad{}\times\sqrt{\frac{|\omega_{in}-\ep||\omega_{out}-\ep|\ep^2}
{v^4\rho_{in}\rho_{out}\rho_{in}'\rho_{out}'}}\,e^{i\Phi_{in}+i\Phi_{out}}\nonumber\\
&&\qquad{}\times\cos\frac{\phi_{in}-\phi_{out}}{2}
\sin\frac{2\varphi_{out}+\phi_{out}'-\phi_{out}}{2}\nonumber\\
&&\qquad{}\times\cos\frac{\phi_{in}'-\phi_{out}'}{2}
\sin\frac{2\varphi_{in}+\phi_{in}'-\phi_{in}}{2},\\
&&\rho_a=\sqrt{x_a^2+(y_a-y_e)^2},\\ &&\rho_a'=\sqrt{(X-x_a)^2+(y_a-Y)^2},\\
&&\phi_a=\arctan\frac{y_a-y_e}{x_a},\quad
\phi_a'=\arctan\frac{y_a-Y}{X-x_a},\\
&&\Phi_a=\frac{|\ep|+i\gamma}{v}\,\rho_a+\frac{|\omega_a-\ep|+i\gamma}{v}\,\rho_a',
\end{eqnarray}\end{subequations}
where $a$ is either ``$in$'' or ``$out$''.
It is convenient to use the paraxial approximation with respect
to the axis connecting the points $(0,y_e)$ and $(X,Y)$.
In this approximation we expand
\begin{subequations}\begin{eqnarray}
&&\rho_a+\rho_a'\approx\frac{X}{\cos\phi_0}
+\frac{X\cos^3\phi_0}{x_a(X-x_a)}\,\frac{(y_a-y_{a0})^2}2,\\
&&\phi_a\approx\phi_0+\frac{y_a-y_{a0}}{x_a}\,\cos^2\phi_0,\\
&&\phi_a'\approx-\phi_0+\frac{y_a-y_{a0}}{X-x_a}\,\cos^2\phi_0,\\
&&y_{a0}=x_a\tan\phi_0,\quad\tan\phi_0=\frac{Y-y_e}{X}.
\end{eqnarray}\end{subequations}
Integrating over $\vec{r}_{in},\vec{r}_{out}$ in the usual way, we obtain
\begin{eqnarray}
\mathcal{M}&=&
-ie^2\sqrt{\frac{\lambda_K}{2}}
\int\limits_{-d_e/2}^{d_e/2}dy_e\,\frac{\sin[\omph{X}/(2v\cos\phi_0)]}{\omph{X}/(2v\cos\phi_0)}\nonumber\\
&&{}\times\exp\left[\left(i\,\frac{\omega_{in}+\omega_{out}}{2}-2\gamma\right)\frac{X}{v\cos\phi_0}\right]\nonumber\\
&&{}\times\sin(\varphi_{in}-\phi_0)\sin(\varphi_{out}-\phi_0).
\end{eqnarray}
This integral can be calculated analogously to the standard
diffraction integral in optics.\cite{BornWolf}
According to Eq.~(\ref{dId2total=}), the integrated intensity into the
full solid angle $4\pi$, summed over two polarizations of the emitted
photon, and integrated over~$\vec{R}$, is given by
\begin{subequations}\begin{eqnarray}
I_D&=&\frac{8\lambda_K}{3}\left(\frac{e^2}c\right)^2\frac{v^2}{c^2}\,
\frac{v}{\omega_{in}L_x}\,\frac{d_e}{L_y}\,\frac{\omega_{in}^2}{\omph^2}
\times\nonumber\\&&{}\times
\left\{\sin^2\varphi_{in}\ln\frac{\omph^2+(4\gamma)^2}{(4\gamma)^2}+\cos^2\varphi_{in}
\right.\nonumber\\ &&{}\qquad\times\left.
\left[\frac{\omph^2}{\omega_{in}^2}\ln\frac{\omega_{in}}\omph
+\frac{2v}{\omega_{in}d_e}\ln\frac{\omph^2+(4\gamma)^2}{(4\gamma)^2}\right]\right\}.
\nonumber\\ \label{IDfragma=}
\end{eqnarray}
This expression is analogous to Eq.~(\ref{IDregularb=}), weighted by
the factor $d_e/L_y$.
The coefficient at the last term, $\propto\cos^2\varphi_{in}$, determines
the minimum of the intensity at $\varphi_{in}=0$ and
has two contributions: the one corresponding to the infinite edge,
and the one due to the finite size of the segment. The latter one
is dominant unless
$d_e\gtrsim(v/\omega_{in})(\omega_{in}/\omph)^2\approx{50}\:\mbox{nm}$
for $\omega_{in}=2\:\mbox{eV}$.
Still, as long as
$\omega_{in}d_e/v\gg{1}$, the ratio between the intensities for the
parallel and perpendicular polarizations is large.
If we use
Eq.~(\ref{dId2phiout=}) for the intensity in a solid angle $o_{out}$ in
the presence of an analyzer, we obtain
\begin{eqnarray}
I_D&=&4\lambda_K\,\frac{o_{out}}{4\pi}
\left(\frac{e^2}c\right)^2\frac{v^2}{c^2}\,
\frac{v}{\omega_{in}L_x}\,\frac{d_e}{L_y}\,\frac{\omega_{in}^2}{\omph^2}
\times\nonumber\\&&{}\times
\left\{\sin^2\varphi_{in}\sin^2\varphi_{out}\ln\frac{\omph^2+(4\gamma)^2}{(4\gamma)^2}
\right. +\nonumber\\ &&{}\qquad+ \left.
\left[\sin^2(\varphi_{in}+\varphi_{out})+\frac{1}2\sin{2}\varphi_{in}\sin{2}\varphi_{out}\right]
\right.\nonumber\\ &&{}\qquad\;\;\times\left.
\frac{v}{\omega_{in}d_e}\ln\frac{\omph^2+(4\gamma)^2}{(4\gamma)^2}
\right. +\nonumber\\ &&{}\qquad+ \left.
\cos^2\varphi_{in}\cos^2\varphi_{out}\right.\nonumber\\ &&{}\qquad\;\;\times\left.
\left[\frac{\omph^2}{\omega_{in}^2}\ln\frac{\omega_{in}}\omph+
\frac{v}{\omega_{in}d_e}\ln\frac{\omph^2+(4\gamma)^2}{(4\gamma)^2}\right]\right\}.\nonumber\\
\label{IDfragmb=}
\end{eqnarray}
\end{subequations}

Eqs. (\ref{IDfragma=}), (\ref{IDfragmb=}) describe the contribution
of a single armchair segment to the Raman intensity. To obtain the
contribution of the whole edge, it is sufficient to multiply these
expression by the total number of such segments and replace $d_e$
by its average, if all segments have the same orientation. It is
crucial, however, that up to three different orientations of armchair
segments are possible, at the angle $\pi/3$ to each other, as
discussed by the author and coworkers in
Ref.~\onlinecite{Casiraghi2009}.

Let us first consider a measurement in the absence of an analyzer.
Since the intensity is a biliniear form of the polarization
vector~$\vec{e}_{in}$, it can always be written in the form
\begin{eqnarray}\label{IDphigeneric=}
I(\varphi_{in})&\propto&\cos^2(\varphi_{in}-\varphi_{max})
+\varepsilon\sin^2(\varphi_{in}-\varphi_{max})\nonumber\\
&=&\frac{1+\varepsilon}{2}+
\frac{1-\varepsilon}{2}\,\cos(2\varphi_{in}-2\varphi_{max}),
\end{eqnarray}
where $\varphi_{max}$ is the angle where the intensity is maximum
and $\varepsilon$ is the ratio between the intensities in the minimum
and in the maximum.
Eq.~(\ref{IDfragma=}) corresponds to $\varphi_{max}=\pi/2$ and
small $\varepsilon\ll{1}$ due to the quantum diffraction.

Let the edge have $N_0$ armchair segments oriented along
the $y$~direction, such as the one considered above, and $N_\pm$
segments oriented at $\pm\pi/3$ to the $y$ axis. Note that the average
direction of the edge may still be arbitrary, as it depends on the
distribution of zigzag segments too. Let each segment be charaterized
by the same values of $\varepsilon$ and $\varphi_{max}$, when the
latter is measured with respect to the corresponding normal. Adding the
contributions, we again obtain an expression of the form of
Eq.~(\ref{IDphigeneric=}), but with different values of parameters:
\begin{subequations}\begin{eqnarray}
&&N_0I(\varphi_{in})+N_+I(\varphi_{in}-\pi/3)+N_-I(\varphi_{in}+\pi/3)\nonumber\\
&&\propto\cos^2(\varphi_{in}-\tilde\varphi_{max})
+\tilde\varepsilon\sin^2(\varphi_{in}-\tilde\varphi_{max}),\\
&&\tilde\varphi_{max}=\varphi_{max}
+\frac{1}2\arctan\frac{\sqrt{3}(N_+-N_-)}{2N_0-N_+-N_-},\label{tvep=}\\
&&\tilde\varepsilon=
\frac{(1+\varepsilon)N_{tot}-(1-\varepsilon)\tilde{N}}%
{(1+\varepsilon)N_{tot}+(1-\varepsilon)\tilde{N}},\\
&&\tilde{N}\equiv\sqrt{N_{tot}^2-3(N_+N_-+N_0N_++N_0N_-)},\\
&&N_{tot}\equiv N_0+N_++N_-.
\end{eqnarray}\end{subequations}
Inspection of Eq.~(\ref{tvep=}) shows that $\tilde\varepsilon\ll{1}$
if and only if (i)~$\vep\ll{1}$ and (ii)~$N_{tot}-\tilde{N}\ll{N}_{tot}$.
The latter condition is equivalent to having one of $N_0,N_+,N_-$ to
be much larger than the others. If these conditions hold, we can write
(assuming that $N_0\gg{N}_+,N_-$ for concreteness)
\begin{equation}
\tilde\varepsilon\approx\varepsilon+\frac{3}{4}\frac{N_++N_-}{N_0}.
\end{equation}
In the opposite case $N_0=N_+=N_-$ we have $\tilde\varepsilon=1$,
so that isotropy is fully restored and no signatures of quantum
diffraction are left.

An analogous summation can be performed in the presence of an analyzer
in the general case,
but the final expressions are very bulky and not very informative. The
qualitative conclusion is the same: the terms which were small compared
to the leading term $\sin^2\varphi_{in}\sin^2\varphi_{out}$ grow as one
adds segments with different orientations. At $N_0=N_+=N_-$ the isotropy
is restored,
\begin{eqnarray}
&&\sin^2\varphi_{in}\sin^2\varphi_{out}
+\sin^2(\varphi_{in}-\pi/3)\sin^2(\varphi_{out}-\pi/3)\nonumber\\
&&{}+\sin^2(\varphi_{in}+\pi/3)\sin^2(\varphi_{out}+\pi/3)\nonumber\\
&&{}=\frac{3}8\sin^2(\varphi_{in}-\varphi_{out})+\frac{9}8\cos^2(\varphi_{in}-\varphi_{out}),
\end{eqnarray}
and signatures of the quantum diffraction are lost.

Let us focus on the special case when the average direction of edge is
zigzag, and it is symmetric on the average, $N_+=N_0$, $N_-=0$.
Then we have
\begin{eqnarray}
I_D&\propto&\sin^2\varphi_{in}\sin^2\varphi_{out}\nonumber\\
&&{}+\sin^2\!\left(\varphi_{in}-\frac\pi{3}\right)
\sin^2\!\left(\varphi_{out}-\frac\pi{3}\right)+O(\vep)\nonumber\\
&=&\frac{3}8+\frac{3}4\cos^2(\varphi_{in}-\varphi_{out})\nonumber\\
&&{}-\cos^2\!\left(\varphi_{in}-\frac\pi{6}\right)
\cos^2\!\left(\varphi_{out}-\frac\pi{6}\right)+O(\vep).\nonumber\\
\label{IDzigzag=}
\end{eqnarray}
The maximum of intensity is reached when both polarizations are
along the average direction of edge. For the unpolarized detection,
we add the contributions with $\varphi_{out}=0$ and
$\varphi_{out}=\pi/2$ and obtain the ratio between the minimum and
the maximum intensity to be $1/3+O(\vep)$; if an analyzer is used
and $\varphi_{in}=\varphi_{out}$, we obtain
$\tilde\varepsilon=1/9+O(\varepsilon)$. These findings agree
with the available experimental data.\cite{Gupta2009,Casiraghi2009}

The dependence of Eq.~(\ref{IDzigzag=}) has a remarkable property:
at $\varphi_{in}=0$, $\varphi_{out}=\pi/3$ (or vice versa) the leading
term vanishes and $I_D=O(\vep)$, i.~e. the quantum limit is still
accessible. In fact, the same will be true for any edge with
only two orientations of the segments (i.~e, for $N_-=0$, but
$N_0\neq{N}_+$, generally speaking).
The ratio of intensity in this minimum to the maximum without
analyzer is given by
\begin{subequations}\begin{eqnarray}
&&\tilde{\vep}=\frac{2}{Z}\left[\frac{v}{\omega_{in}d_e}+
\frac{\omph^2}{4\omega_{in}^2}\,\ln\frac\omph{\omega_{in}}
\left(\ln\frac{\omph^2+(4\gamma)^2}{(4\gamma)^2}\right)^{-1}\right],
\nonumber\\ && \label{IDquantum=}\\
&&Z\equiv{1}+\frac{\sqrt{N_0^2+N_+^2-N_0N_+}}{N_0+N_+}.
\end{eqnarray}\end{subequations}

\section{Conclusions}

We have studied scattering of Dirac electrons on a graphene edge.
For a translationally invariant edge (such
as zigzag or armchair or another edge with a certain spatial
period) the reflection can be described by an effective
low-energy boundary condition for the electronic wave
function.\cite{McCann2004,Akhmerov2008}
For edges which are rough on the atomic scale we have proposed a
random-matrix model which
describes random scattering of electrons on the edge, respecting
the particle conservation and time-reversal symmetry. Essentially,
each point of the edge acts as an independent point scatterer with
a random rotation of the valley state. We have also considered
edges consisting of zigzag and armchair distinct segments longer
than the electron wavelength, each segment can be treated as a
piece of an ideal edge, while the small corrections due to quantum
diffraction, can be found using the Huygens-Fresnel principle for
Dirac electrons, analogously to the standard treatment of diffraction
in the classical optics.

Next, we have calculated the intensity of the edge-induced $D$~peak
in the Raman scattering spectrum of graphene. It is shown how the
quasiclassical character of the electron motion manifests itself in
the polarization dependence of the intensity. For an ideal armchair
edge the maximum of intensity corresponds to the case when both the
polarizer and the analyzer are along the edge, and the large ratio of
intensities in the maximum and the minimum turns out to be determined
by the quantum corrections to the quasiclassical motion of the
photoexcited electron and the hole. For an edge consisting of randomly
distributed zigzag and armchair segments of the length significantly
exceeding the electron wavelength, the effect of quantum diffraction
can be masked by the presence of armchair segments of different
orientations. The maximum and the minimum of the intensity are
determined by the number of the armchair segments with different
orientations, rather than the average direction of the edge. If only
two orientations of armchair segments are present in the edge, the
quantum diffraction limit can still be probed by a careful choce of
the polarizer and the analyzer (the polarizer should be oriented
perpendicularly to one of the armchair directions, the analyzer
perpendicularly to the other one). For an edge, rough at the atomic
scale, no segments can be identified, and the intensity reaches its
maximum for the polarization along the average direction of the edge.
The ratio of the maximum and the minimum intensity is determined by
the ability of the edge to scatter electrons at oblique angles.

As the whole Raman process is edge-assisted, one can pose the question
about the characteristic length scale which restricts the process to
the vicinity of the edge. We find that the answer is not unique, and
the length scale depends on the specific observable under study.
If one is interested in the total intensity or its polarization
dependence, the effective length scale is $v/\omph$ ($v$~being the
electronic velocity and $\omph$~the phonon frequency). However, if one
makes a spatially resolved experiment, measuring the dependence of
the intensity on the position of the excitation spot, the relevant
length scale is the electron inelastic scattering length $v/(2\gamma)$.
We have thus found a qualitative agreement with the interpretation of
Ref.~\onlinecite{Novotny}, but we argued the inelastic scattering
length found in that work is too large to be consistent with other
available information on electron inelastic scattering in graphene.

\section{Acknowledgements}

The author is grateful to A. C. Ferrari, S. Piscanec, and M. M. Fogler
for stimulating discussions.

\appendix

\section{Scattering matrix for an edge}\label{app:Smatrix}

The representation of the 4-column vector~$\psi$, natural for scattering
problems, is that of Ref.~\onlinecite{AleinerEfetov},
Eq.~(\ref{Aleinerrep=}).
For 4-column vectors which can be represented as a direct product
\begin{equation}
\left[\begin{array}{c}
x_1y_1 \\ x_2y_1 \\ x_1y_2 \\ x_2y_2
\end{array}\right]\equiv
\left[\begin{array}{c} x_1 \\ x_2 \end{array}\right]
\otimes\left[\begin{array}{c} y_1 \\ y_2 \end{array}\right]
\equiv
\left[\begin{array}{c} x_1 \\ x_2 \end{array}\right]
\otimes(y_1\phi_K+y_2\phi_{K'}),
\end{equation}
the $\Sigma$ ~matrices act on the $x$~variables, while the
$\Lambda$~matrices act on the $y$~variables. The basis in the
valley subspace is denoted by $\phi_K,\phi_{K'}$ for future
convenience.

To keep the formulas compact we assume that the edge is along the
$y$~direction and graphene is occupying the half-space $x>0$, i.~e.,
$\vec{n}=\vec{e}_x$.
The average orientation of a disordered edge with $d_e\sim{a}$
does not have to be correlated with any cristallographic direction,
so the description is equivalent for all orientations.
Thus, expressions for a general orientation of the edge are obtained
by replacing the coordinates by
$x\to\vec{n}\cdot\vec{r}$, $y\to[\vec{n}\times\vec{r}]_z$,
and the polar angles by $\varphi\to\varphi+\varphi_\vec{n}$.

Let us label the eigenstates of the problem by three quantum numbers:
(i)~$\ep$, the energy of the electron;
(ii)~$p_y$, the $y$-component of the momentum of the incident plane wave
(obviously, $|p_y|<|\ep|/v$),
(iii)~$\valley$, the valley index (we write $\kappa=\pm{1}$ for
$K$~and~$K'$, respectively).
The most general wave function of such an eigenstate has the form
\begin{subequations}\begin{eqnarray}
&&\psi_{\ep,{p}_y,\valley}(x,y)=
\frac{e^{ip_yy-ip_xx}}{\sqrt{L_xL_y}}\,
\sqrt{\frac{|\ep|}{vp_x}}\,\overleftarrow\psi_{\ep,{p}_y,\valley}+
\nonumber\\ && \qquad\qquad{}+
\sum_{p_y'}S_{p_y'p_y}^{\,\valley'\valley}\,
\frac{e^{ip_y'y+ip_x'x}}{\sqrt{L_xL_y}}\,
\sqrt{\frac{|\ep|}{vp_x'}}\,\overrightarrow\psi_{\ep,{p}_y',\valley'},
\label{scatteringsolution=}\\
&&p_x\equiv\sqrt{(\ep/v)^2-p_y^2},\quad p_x'\equiv\sqrt{(\ep/v)^2-(p_y')^2},\\
&&\sum_{p_y'}\equiv\int\limits_{-|\ep|/v}^{|\ep|/v}\frac{L_y\,dp_y'}{2\pi},\quad
\delta_{p_yp_y'}=\frac{2\pi}{L_y}\,\delta(p_y-p_y')\\
&&\overleftarrow\psi_{\ep,{p}_y,\valley}=
\frac{1}{\sqrt{2}}\left[\begin{array}{c}
e^{i(\varphi_{p_y}-\pi)/2} \\ e^{i(\pi-\varphi_{p_y})/2}\sign\ep
\end{array}\right]\otimes\phi_\valley,\\
&&\overrightarrow\psi_{\ep,{p}_y,\valley}=
\Sigma_y\overleftarrow\psi_{\ep,{p}_y,\valley}=
\frac{1}{\sqrt{2}}\left[\begin{array}{c}
e^{-i\varphi_{p_y}/2}\sign\ep \\ e^{i\varphi_{p_y}/2}
\end{array}\right]\otimes\phi_\valley,\\
&&\varphi_{p_y}=\arctan\frac{p_y}{p_x},\quad -\frac\pi{2}\leq\varphi_{p_y}\leq\frac\pi{2}.
\end{eqnarray}\end{subequations}
The prefactor $\sqrt{|\ep|/(vp_x)}$ in Eq.~(\ref{scatteringsolution=}) fixes
the component of the probability current perpendicular to the boundary
to be $j_x=-v\sign\ep/(L_xL_y)$ for the incident wave.\cite{holes}
Finally, ${S}_{p_y'p_y}^{\,\valley'\valley}$ are the
scattering matrix elements, which may depend on~$\ep$.
Note that $\varphi_{p_y}=\varphi_{p_y'}$ corresponds to the specular
reflection. The wave functions $\psi_{\ep,{p}_y,\valley}(x,y)$ are
orthogonal,
\begin{eqnarray}
&&\int\limits_{-\infty}^\infty{d}y\int\limits_0^\infty{dx}\,
\psi^\dagger_{\ep,p_y,\valley}(x,y)\,
\psi_{\ep'\!,p_y',\valley'}(x,y)=\nonumber\\
&&=\frac{2\pi{v}}{L_x}\,\delta(\ep-\ep')\,
\delta_{p_yp_y'}\,\delta_{\valley\valley'},
\end{eqnarray}
provided that the scattering matrix is unitary:
\begin{equation}
\sum_{p_y',\valley'}\left({S}^{\,\valley'\valley}_{p_y'p_y}\right)^*
{S}^{\,\valley'\valley''}_{p_y'p_y''}
=\delta_{\valley\valley''}\,\delta_{p_yp_y''}. \label{unitaritySpp=}
\end{equation}
The same condition ensures the conservation of the flux, i.~e.
that the current $j_x=v\sign\ep/(L_xL_y)$ for the scattered part
of the wave function~(\ref{scatteringsolution=}).
Reflection from
a regular edge with wave function given by Eq.~(\ref{reflectingsolution=}),
corresponds to
${S}^{{\valley'\valley}}_{p_y'p_y}=%
\delta_{p_y'p_y}\phi_{\valley'}^\dagger{S}_\Lambda\phi_\valley$.

The time reversal symmetry imposes another condition on the
scattering matrix (reciprocity condition).
Namely, the time-reversed wave function
$U_t\psi_{\ep,{p}_y,\valley}^*(x,y)$ describes an eigenstate
of the problem, and thus it must be a linear combination of
$\psi_{\ep,{p}_y',\valley'}(x,y)$ with different $p_y',\valley'$.
Noting that
\begin{subequations}\begin{eqnarray}
&&U_t\overleftarrow\psi^*_{\ep,-p_y,-\valley}=
i\valley\overrightarrow\psi_{\ep,p_y,\valley},
\label{trevleft=}\\
&&U_t\overrightarrow{\psi}^*_{\ep,-p_y,-\valley}=
-i\valley\overleftarrow\psi_{\ep,p_y,\valley}.
\label{trevright=}
\end{eqnarray}\end{subequations}
we obtain the reciprocity condition,
\begin{equation}\label{reciprocity=}
S^{\,\valley'\valley}_{p_y'p_y}=-\valley\valley'\,S^{-\valley,-\valley'}_{-p_y,-p_y'},
\end{equation}
which can also be written in the matrix form as
$S_{p_y'p_y}=-\Lambda_y\,S^T_{-p_y,-p_y'}\Lambda_y$, where $S^T$
denotes the $2\times{2}$ matrix transpose.

To relate the scattering matrix to the $T$-matrix in
Eq.~(\ref{wfTmatrix=}), we compare the $x\gg{v}/|\ep|$ asymptotics of
the corresponding wave functions. The Green's function
$G_0(\vec{r}-\vec{r}_e,\ep)$ entering Eq.~(\ref{wfTmatrix=}) can be
represented as follows:
\begin{equation}\begin{split}
&G_0(x,y-y_e,\ep)=\\
&=\int\limits_{-\infty}^{\infty}\frac{d\ep'}{2\pi{v}}
\int\limits_{-|\ep'|/v}^{|\ep'|/v}\frac{dp_y}{2\pi}\,
\frac{|\ep'|}{vp_x}\,\frac{e^{ip_y(y-y_e)}}{\ep-\ep'+i0^+\sign\ep}\\
&{}\times\sum_\valley\left(
\overrightarrow\psi_{\ep',{p}_y,\valley}\overrightarrow\psi_{\ep',{p}_y,\valley}^\dagger
e^{ip_xx}
+\overleftarrow\psi_{\ep',{p}_y,\valley}\overleftarrow\psi_{\ep',{p}_y,\valley}^\dagger
e^{-ip_xx}\right),
\end{split}\end{equation}
where we suppressed the damping $\gamma_\ep\to{0}^+$
and denoted $p_x\equiv\sqrt{(\ep'/v)^2-p_y^2}$).
To determine the $x\gg{v}/|\ep|$ asymptotics of the Green's function,
we shift the $\ep'$~integration contour in the upper (lower) complex
half-plane at $\ep'>0$ ($\ep'<0$) for the term $\propto{e}^{ip_xx}$,
and in the lower (upper) half-plane at $\ep'>0$ ($\ep'<0$) for the term
$\propto{e}^{-ip_xx}$.  The magnitude $\ep''$ of the shift is such that
$v/x\ll\ep''\ll|\ep'|$ As a result, (i)~the contribution from the term
$\propto{e}^{-ip_xx}$ will be exponentially small except for the small
region near zero, $|\ep'|\sim{v}/x$; (ii)~the contribution from the term
$\propto{e}^{ip_xx}$ will be determined by the pole $\ep'=\ep+i0^+\sign\ep$,
the rest of the contour contributing an exponentially small quantity;
(iii)~the contribution from $|\ep'|\sim{v}/x$ will be small as $v/(|\ep|x)$.
Thus, at $x\gg{v}/|\ep|$ we arrive at
\begin{eqnarray}
G_0(x,y-y_e,\ep)&=&
\frac{i}{v}\int\limits_{-|\ep|/v}^{|\ep|/v}\frac{dp_y}{2\pi}\,
\frac{\ep}{vp_x}\,{e}^{ip_y(y-y_e)+ip_xx}\nonumber\\
&&{}\times\sum_\valley
\overrightarrow\psi_{\ep,{p}_y,\valley}
\overrightarrow\psi_{\ep,{p}_y,\valley}^\dagger.
\end{eqnarray}
Substituting this expression and the first term of Eq.~(\ref{reflectingsolution=})
into Eq.~(\ref{gfTmatrix=}), and comparing the result with last term of
Eq.~(\ref{reflectingsolution=}), we obtain the relation
\begin{eqnarray}
S_{p_y'p_y}^{\,\valley'\valley}&=&
\frac{i\sign\ep}{v}\int\frac{dy_e}{L_y}\,e^{-i(p_y'-p_y)y_e}\nonumber\\
&&{}\times\frac{\overrightarrow\psi_{\ep,{p}_y',\valley'}^\dagger
T(\varphi_{p_y'},\pi-\varphi_{p_y};y_e)
\overleftarrow\psi_{\ep,{p}_y,\valley}}%
{\sqrt{\cos\varphi_{p_y'}\,\cos\varphi_{p_y}}}.
\end{eqnarray}
Here $\varphi_{p_y'}$ is the polar angle of the direction~$\vec{s}$,
and $\pi-\varphi_{p_y}$ is the polar angle of the incident
momentum~$\vec{p}$, directed towards the edge.
Using Eqs.~(\ref{trevleft=}),~(\ref{trevright=}), one can establish the
equivalence between the reciprocity condition,
Eq.~(\ref{reciprocity=}), and Eq.~(\ref{Ttimerev=}). Using the
unitarity condition, Eq.~(\ref{unitaritySpp=}), one fixes the prefactor
$\pi{v}/|\ep|$ in front of the $\delta$-function in
Eq.~(\ref{COEaverage=}).

\section{Integrals of Sec.~\ref{sec:integrated}}\label{app:integrals}

First, let us focus on Eq.~(\ref{Minitial=}) for $\mathcal{M}_\|$. Upon
integration over $y_{in},y_{out}$, it takes the following form:
\begin{subequations}
\begin{eqnarray}
\mathcal{M}_\|&=&\frac{e^2}{\omega_{in}}\,
\sqrt{\frac{4\lambda_Kv}{i\pi}}\int\limits_{-\infty}^\infty\frac{d\ep}v
\int\limits_0^Xdx_{in}\,dx_{out}\,e^{i\Phi_x-2\gamma{X}/v}\nonumber\\
&&{}\times\left(\frac{x_{in}+x_{out}}{|\ep|}
+\frac{X-x_{in}}{|\omega_{in}-\ep|}
+\frac{X-x_{out}}{|\omega_{out}-\ep|}\right)^{-1/2},\\
\Phi_x&=&\frac{|\ep|}v\,(x_{in}+x_{out})+
\frac{|\omega_{in}-\ep|}v\,(X-x_{in})\nonumber\\
&&{}+\frac{|\omega_{out}-\ep|}v\,(X-x_{out}).
\end{eqnarray}
\end{subequations}
If one neglects the square root and integrates first over $x_{in}$, the
energy is constrained to be $\ep\approx\omega_{in}/2$ with the
precision $\sim{v}/X$. If one integrates over $x_{out}$, then
$\ep\approx\omega_{out}/2$ (which is not inconsistent with the previous
condition as long as $X\sim{v}/\omph$). If one also removes the moduli
and integrates over $\ep$, then $x_{in}+x_{out}\approx{X}$.

To implement these observations more rigorously, let us introduce new
variables,
\begin{subequations}\begin{eqnarray}
&&\tilde{x}=\frac{x_{in}+x_{out}-X}2,\quad x_0=x_{in}-x_{out},\\
&&\tilde\ep=\ep-\frac{\omega_{in}+\omega_{out}}4,\\
&&\Phi_x=4\,\frac{\tilde\ep\tilde{x}}{v}-\frac{\omph{x}_0}{2v}
+\frac{\omega_{in}+\omega_{out}}{2v}\,X.
\end{eqnarray}\end{subequations}
Recalling that the typical scale of the $\tilde\ep$~dependence of the
rest of the integrand is $\omega_{in}$, and that of the
$\tilde{x}$~dependence is~$X$, we rewrite
\begin{equation}
4\,\frac{\tilde\ep\tilde{x}}{v}= \frac{\omega_{in}X}{v}
\left(\frac{\tilde\ep}{\omega_{in}}+\frac{\tilde{x}}X\right)^2
-\frac{\omega_{in}X}{v}
\left(\frac{\tilde\ep}{\omega_{in}}-\frac{\tilde{x}}X\right)^2.
\end{equation}
Taking the two expressions in the brackets as new integration
variables, we can use the stationary phase approximation justified by
$\omega_{in}X/v\gg{1}$. As a result, the integral is contributed by
$|\tilde\ep|\sim\sqrt{\omega_{in}v/X}\ll\omega_{in}$ and by
$|\tilde{x}|\sim\sqrt{Xv/\omega_{in}}\ll{X}$. Thus, in the rest of the
integrand they can be simply set to zero, i.~e. we approximate
$e^{4i\tilde\ep\tilde{x}/v}\approx%
(\pi{v}/2)\delta(\tilde\ep)\delta(\tilde{x})$.
The remaining integration over $x_0$ is elementary:
\begin{eqnarray}\nonumber
&&\int\limits_{-X}^Xdx_0\int\limits_{-(X-|x_0|)/2}^{(X-|x_0|)/2}d\tilde{x}
\int\limits_{|\tilde\ep|\lesssim\omega_{in}/2}\frac{d\tilde\ep}v\,
e^{i\Phi_x}\\ &&\approx\pi{e}^{i(\omega_{in}+\omega_{out})X/(2v)}\,
\frac{\sin[\omph{X}/(2v)]}{\omph/(2v)}.
\end{eqnarray}

For the integral in Eqs.~(\ref{Mprimea=}),~(\ref{Mprimeb=}) we have:
\begin{subequations}\begin{eqnarray}
&&\tilde{x}=\frac{x_{in}-x_{out}-X}2,\quad x_0=\frac{x_{in}+x_{out}-X}2,\\
&&\tilde\ep=\ep-\frac{\omega_{in}+\omega_{out}}4,\\
&&\Phi_x'=-2\,\frac{\tilde\ep\tilde{x}}{v}+\frac{\omph{x}_0}{v}
+\frac{\omega_{in}X}v.
\end{eqnarray}\end{subequations}
\begin{eqnarray}
&&\int\limits_{|\tilde\ep|\lesssim\omega_{in}/2}\frac{d\tilde\ep}v\,
\int\limits_0^Xd\tilde{x}\int\limits_{-\tilde{x}/2}^{\tilde{x}/2}dx_0\,
e^{i\Phi_x'}\nonumber\\ 
&&=e^{i\omega_{in}X/v}\int\limits_{|\tilde\ep|\lesssim\omega_{in}/2}
\frac{v\,d\tilde\ep}{2\omph}\nonumber\\
&&\times\left[\frac{e^{-i(2\tilde\ep-\omph/2)X/v}-1}{2\tilde\ep-\omph/2}
-(\omph\to-\omph)\right].
\end{eqnarray}
If we extend the limits of the $\tilde\ep$~integration to infinity,
the integral vanishes by analyticity. Thus, it is determined by
the region $|\ep|\sim\omega_{in}$ and can be estimated as
$\sim{v}/\omega_{in}$.

Now let us pass to the derivation of Eq.~(\ref{Mperpfinal=}) for
$\mathcal{M}_\perp$. The difference from the previous case is that
integration over $y_{in},y_{out}$ in the derivation is more subtle.
Namely, we encounter oscillating integrals of the kind
$\int{y}^2e^{i\alpha{y}^2}dy$. This integral is assigned the value
$\sqrt{\pi/4}\,(-i\alpha)^{-3/2}$, which may be understood as the
analytical continuation from the upper complex half-plane of~$\alpha$.
One may argue that since the integral is divergent for real~$\alpha$,
it is determined by $|y|\sim{1}/\sqrt{\Im\alpha}$ which are not small,
so the expansion to $y^2$ both in the exponential and in the
pre-exponential factor is not valid.

\begin{figure}
\includegraphics[width=8cm]{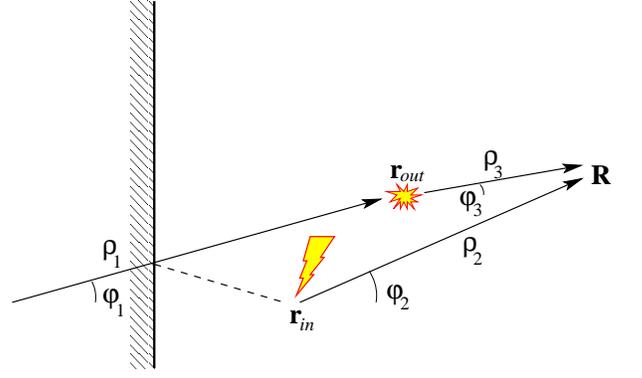}
\caption{\label{fig:trajectmess}(Color on-line.) Geometry corresponding
to Eq.~(\ref{Mnonparaxial=}). The solid black arrows show the vectors
$\vec\rho_1,\vec\rho_2,\vec\rho_3$.}
\end{figure}

Let us write down the expression for the matrix element without using
the paraxial approximation [we omit the $1/\zeta$ terms of the
expansion in Eq.~(\ref{GreenfLarger=}) as they do not produce $y^2$ terms
and do not represent any convergence problem],
\begin{eqnarray}
\mathcal{M}&=&
\sqrt{\frac{i\lambda_K}{\pi^3v}}\,\frac{e^2}{\omega_{in}}
\int\limits_{-\infty}^\infty\frac{d\ep}v \int
d^2\vec{r}_{in}\,d^2\vec{r}_{out}\,
\cos\frac{\varphi_2-\varphi_3}{2}\nonumber\\
&&{}\times\sin\frac{2\varphi_{in}+\varphi_1-\varphi_2}{2}
\sin\frac{2\varphi_{out}-\varphi_1-\varphi_3}{2}\nonumber\\
&&{}\times{e}^{(i|\ep|-\gamma)\rho_1/v+(i|\omega_{in}-\ep|-\gamma)\rho_2/v
+(i|\omega_{out}-\ep|-\gamma)\rho_3/v}\nonumber\\
&&{}\times\sqrt{\frac{|(\omega_{in}-\ep)(\omega_{out}-\ep)\ep|}
{\rho_1\rho_2\rho_3}}.\label{Mnonparaxial=}
\end{eqnarray}
The distances $\rho_{1,2,3}$ and the angles $\varphi_{1,2,3}$ are shown
in Fig.~\ref{fig:trajectmess}. Again, integration over $\ep/v$ gives
effectively $\delta(\rho_1-\rho_2-\rho_3)$, so the exponential in the
integrand becomes
$e^{(i\omega_{in}-2\gamma)\rho_2/v+(i\omega_{out}-2\gamma)\rho_3/v}$.
The phase of this exponential is rapidly oscillating, and the
stationary phase condition gives precisely $y_{in},y_{out}\to{0}$. The
straightforward application of the stationary phase method for
$\varphi_{in}=\varphi_{out}=0$ is impeded by the presence of two sine
functions in front of the exponential.

First of all, let us check the convergence of the integral at large
distances, i.~e. assuming $r_{in},r_{out}\gg{X}$. Then it is convenient
to use the polar coordinates for $\vec{r}_{in},\vec{r}_{out}$ whose
polar angles are almost $\pi+\varphi_2,\pi+\varphi_3$. The
$\delta$-function fixes $\varphi_2=\pi-\varphi_1$,
$\varphi_3=\pi+\varphi_1$. We can set $X=0$ everywhere except the
Jacobian:
\begin{subequations}\begin{eqnarray}
&&\rho_1\approx{r}_{in}r_{out}
-\frac{r_{in}r_{out}}{2(r_{in}+r_{out})}\,(\varphi_2+\varphi_3)^2,\\
&&\rho_{2,3}\approx{r}_{in,out}+X\cos\varphi_{2,3},\\
&&\delta(\rho_1-\rho_2-\rho_3)\approx
\sqrt{\frac{r_{in}+r_{out}}{Xr_{in}r_{out}}}\,
\delta(\varphi_2+\varphi_3).
\end{eqnarray}\end{subequations}
This gives a finite value,
\begin{eqnarray}
\mathcal{M}&=&
\sqrt{\frac{i\lambda_K}{\pi^3v}}\,\frac{e^2}{\omega_{in}}
\int\limits_{-\pi/2}^{\pi/2}d\varphi_1 \int
r_{in}\,dr_{in}\,r_{out}\,dr_{out}\nonumber\\
&&{}\times \sqrt{\frac{r_{in}+r_{out}}{Xr_{in}r_{out}}}\,
\sqrt{\frac{(\omega_{in}/2)^3}{(r_{in}+r_{out})r_{in}r_{out}}}\nonumber\\
&&{}\times{e}^{(i\omega_{in}-2\gamma)r_{in}/v
+(i\omega_{out}-2\gamma)r_{out}/v}\nonumber\\
&&{}\times\cos\varphi_1\cos(\varphi_{in}+\varphi_1)
\cos(\varphi_{out}-\varphi_1)\nonumber\\
&\approx&\sqrt{\frac{i\lambda_Ke^4}{8\pi^3}\frac{v}{\omega_{in}X}}\,
\frac{v}{\omega_{in}}\nonumber\\
&&{}\times \left(\frac{2}3\,\sin\varphi_{in}\sin\varphi_{out}
-\frac{4}3\,\cos\varphi_{in}\cos\varphi_{out}\right).
\end{eqnarray}
Note that for any polarization it is smaller than $\mathcal{M}_\|$ from
Eq.~(\ref{Mfinal=}) by a factor $\omph/\omega_{in}$.

Thus, we are facing a situation of the kind
\begin{equation}
F(\alpha)=\int\limits_{-\infty}^\infty f(y)\,e^{i\alpha\Phi(y)}\,dy,
\end{equation}
where (i)~the function $\Phi(y)$ is growing as $|y|$ at $|y|\to\infty$
and has an extremum at $y=0$, where it can be expanded as
$\Phi(y)=\Phi(0)+\Phi''(0)\,y^2/2+O(y^4)$, (ii)~$\alpha$ is large and
is assumed to have a positive imaginary part, and
(iii)~$f(y)=f''(0)\,y^2/2+O(y^3)$ at $y\to{0}$, while the behavior of
the function $f(y)$ at $|y|\to\infty$ is such that the in the limit
$\Im\alpha\to{0}$ the integral $F(\alpha)$ remains finite. Then
$F(\alpha)$ is analytic in the upper complex half-plane of~$\alpha$.
Thus, the expansion
\begin{equation}
F(\alpha)=\frac{\sqrt{\pi/2}\,f''(0)}{[-i\alpha\Phi''(0)]^{3/2}}
+O((-i\alpha)^{-5/2}),
\end{equation}
established for large positive imaginary~$\alpha$, holds in the whole
upper half-plane, including the vicinity of the real axis. The
two-dimensional integration over $y_{in},y_{out}$, leading to
Eq.~(\ref{Mperpfinal=}) is fully analogous.

\section{Detailed expressions for Sec.~\ref{sec:regSpatial}}
\label{app:kernel}

Note that length $l_2$ is not well defined in the general case. Namely,
for an arbitrary kernel $\mathcal{K}(x,x')$ and an arbitrary shape of
$\mathcal{E}_{in}(x)$ the expansion in the derivatives of
$\mathcal{E}_{in}(x)$ does not automatically ``wrap'' into the expansion
in the derivatives of $w(x)$, Eq.~(\ref{expandprofile=}).
However, $l_2$ can be defined in two
particular cases which are most important for us. It is convenient to
introduce an auxiliary notation: for an arbitrary function $\mathcal{F}(x,x')$
we define
\begin{equation}
\langle\langle\mathcal{F}(x,x')\rangle\rangle
\equiv\int\limits_0^\infty\mathcal{F}(x,x')\,\mathcal{K}(x,x')\,
\frac{dx\,dx'}{l_0^2}-\mathcal{F}(l_1,l_1).
\end{equation}
One case when the length $l_2$ can be defined is when the kernel
$\mathcal{K}(x,x')$ is such that
\begin{equation}\label{Kernelproperty=}
\langle\langle{x}^2\rangle\rangle+\langle\langle(x')^2\rangle\rangle
=2\langle\langle{xx'}\rangle\rangle,
\end{equation}
then one can set $l_2=\sqrt{\langle\langle{xx'}\rangle\rangle}$.
For the kernel given by Eq.~(\ref{spatialkernel=}) this property
holds in the limit $\ell_\gamma\gg\ellph$:
\begin{eqnarray}
&&\int\limits_0^\infty{x}^2\mathcal{K}(x,x')\,dx\,dx'=
-2\ellph^4\ln\frac{\ell_\gamma^2+4\ellph^2}{4\ellph^2}\nonumber\\
&&\qquad{}+\ell_\gamma^2\ellph^2\,
\frac{\ell_\gamma^4-4i\ell_\gamma^3\ellph+20\ell_\gamma^2\ellph^2+32\ellph^4}
{4(\ell_\gamma^2+4\ellph^2)^2},\\
&&\int\limits_0^\infty{x}x'\mathcal{K}(x,x')\,dx\,dx'=
\ellph^4\ln\frac{\ell_\gamma^2+4\ellph^2}{4\ellph^2}
\nonumber\\&&\qquad{}
+\frac{\ell_\gamma^2\ellph^2}4\,
\frac{\ell_\gamma^2-4\ellph^2}{\ell_\gamma^2+4\ellph^2},
\end{eqnarray}
The second case when $l_2$ can be defined is the gaussian profile
$\mathcal{E}_{in}(x)\propto{e}^{-(x-x_0)^2/(2L^2)}$,
which has a special property
$\mathcal{E}_{in}''(x)\mathcal{E}_{in}(x)=[\mathcal{E}_{in}'(x)]^2
-[\mathcal{E}_{in}(x)]^2/L^2$. Then the difference between the 
left-hand and the right-hand sides of Eq.~(\ref{Kernelproperty=}) can be
absorbed in the overall coefficient, so that instead of
Eq.~(\ref{spatialkernel=}) we have its slightly modified version:
\begin{subequations}\begin{eqnarray}
&&\mathcal{I}_{x_0}=Al_0^2\left[w(l_1-x_0)+\frac{l_2^2}2\,w''(l_1-x_0)\right]
+O(\ell^3/L^3),\nonumber\\ && \label{expandgaussian=}\\
&&A=1
-\frac{\langle\langle{x}^2\rangle\rangle+\langle\langle(x')^2\rangle\rangle
-2\langle\langle{xx'}\rangle\rangle}{4L^2},\\
&&l_2^2=\frac{\langle\langle{x}^2\rangle\rangle+\langle\langle(x')^2\rangle\rangle
+2\langle\langle{xx'}\rangle\rangle}{4}=\nonumber\\
&&\quad=
\frac{\ell_\gamma^2(\ell_\gamma^4+10\ell_\gamma^2\ellph^2+8\ellph^4)}
{4(\ell_\gamma^2+4\ellph^2)^2}\left(\ln\frac{\ell_\gamma^2+4\ellph^2}{4\ellph^2}\right)^{-1}
\nonumber\\ &&\qquad{}
-\frac{\ell_\gamma^6}{4(\ell_\gamma^2+4\ellph^2)^2}
\left(\ln\frac{\ell_\gamma^2+4\ellph^2}{4\ellph^2}\right)^{-2}
-\frac{\ellph^2}2.
\end{eqnarray}\end{subequations}


\begin{thebibliography}{99}
\bibitem{holes}
For the states with $\ep<0$ the direction of the electron
group velocity is opposite to that of momentum. Thus, for $\ep<0$ the
first term of the wave function~(\ref{scatteringsolution=}) represents
an outgoing flux of electrons, and the second term -- to an incoming
flux. However, we prefer to work with the chronologically ordered
Green's function $G(\vec{p},\ep)$, defined in Eq.~(\ref{Greenfp=}),
rather than for the retarded Green's function $G^R(\vec{p},\ep)$, the
analytical continuation of $G(\vec{p},\ep)$ from the positive semiaxis
of~$\ep$ through the upper complex half-plane of~$\ep$. For this
purpose we work with the solutions~(\ref{scatteringsolution=}) where
the first term of the wave function~(\ref{scatteringsolution=})
represents an incoming flux for electrons if $\ep>0$ and for holes if
$\ep<0$.
\end{thebibliography}
\end{document}